\documentclass[aps,pre,twocolumn,superscriptaddress,10pt]{revtex4-2}
\usepackage[utf8]{inputenc}
\usepackage{color}
\usepackage{amsmath}
\usepackage{amssymb}
\usepackage{xcolor} 
\usepackage{graphicx}
\usepackage{esint}
\usepackage{appendix}
\usepackage{comment}
\usepackage{CJK}    
\usepackage{hyperref}

\makeatletter
\@ifundefined{textcolor}{}
{%
\definecolor{red}{rgb}{0.75, 0.1, 0.1}
 \definecolor{BLACK}{gray}{0}
 \definecolor{WHITE}{gray}{1}
 \definecolor{RED}{rgb}{1,0,0}
 \definecolor{GREEN}{rgb}{0,1,0}
 \definecolor{BLUE}{rgb}{0,0,1}
 \definecolor{CYAN}{cmyk}{1,0,0,0}
 \definecolor{MAGENTA}{cmyk}{0,1,0,0}
 \definecolor{YELLOW}{cmyk}{0,0,1,0}
}

\makeatother

\begin{document}
\begin{CJK*}{UTF8}{gbsn}
\title{
Nonlinear Response Identities and Bounds for Nonequilibrium Steady States
}

\author{Ruicheng Bao}
\email{Corresponding author: ruicheng@g.ecc.u-tokyo.ac.jp}
\affiliation{Department of Physics, Graduate School of Science, 7-3-1,
The University of Tokyo, Hongo, Bunkyo-ku, Tokyo 113-0033, Japan}

\author{Shiling Liang (梁师翎)}
\email{Corresponding author: shiling@pks.mpg.de}
\affiliation{Center for Systems Biology Dresden, 01307 Dresden, Germany}
\affiliation{Max Planck Institute for the Physics of Complex Systems, 01187 Dresden, Germany}
\affiliation{Max Planck Institute of Molecular Cell Biology and Genetics, 01307 Dresden, Germany}
\begin{abstract}
Understanding how systems respond to external perturbations is fundamental to statistical physics. For systems far from equilibrium, a general framework for response remains elusive. While progress has been made on the linear response of nonequilibrium systems, a theory for the nonlinear regime under finite perturbations has been lacking. Here, building on a novel connection between response and mean first-passage times in continuous-time Markov chains, we derive a comprehensive theory for the nonlinear response to archetypal local perturbations. We establish an exact identity that universally connects the nonlinear response of any observable to its linear counterpart via a simple scaling factor. This identity directly yields universal bounds on the response magnitude. Furthermore, we establish a universal bound on response resolution -- an inequality constraining an observable's change by its intrinsic fluctuations -- thereby setting a fundamental limit on signal-to-noise ratio. These results provide a rigorous and general framework for analyzing nonlinear response far from equilibrium, which we illustrate with an application to transcriptional regulation.
\end{abstract}
\maketitle
\end{CJK*}  

\textit{Introduction.}---A cornerstone of statistical physics is the quantitative description of a system's response to external perturbations. Near thermal equilibrium, this challenge is elegantly met by the fluctuation-dissipation theorem (FDT), which establishes a universal link between the response of an observable and its spontaneous correlation functions \cite{kubo1966Fluctuationdissipation}. This powerful framework, however, breaks down for the vast class of systems operating far from equilibrium. Such systems are ubiquitous, particularly in biology, where nonequilibrium dynamics are essential for functions like biochemical sensing, adaptation, and regulation \cite{cao2025Stochastic,tenwolde2016Fundamental,qian2007Phosphorylation,ge2012Stochastic}. The absence of a general and equally powerful response theory for these far-from-equilibrium systems remains a central, unresolved problem in statistical physics.

Recent advances in stochastic thermodynamics have established a powerful framework for analyzing nonequilibrium systems at mesoscopic scale \cite{peliti2021stochastic}. On the relation between fluctuation and response, general inequalities for both stationary and non-stationary cases were derived in \cite{dechant2020Fluctuation} using stochastic trajectories and information theory. A parallel route uses algebraic graph theory to build thermodynamic bounds on steady-state responses \cite{owen2020Universal,owen2023Thermodynamic,owen2023Size,fernandesmartins2023Topologically}. More recently, identities and thermodynamic bounds were developed for response in nonequilibrium steady states, ranging from responses of steady-state distribution and currents to general observables \cite{maes2020Response,malloryigoshin2020,aslyamov2024General,aslyamov2024Nonequilibrium,ptaszynski2024Dissipation,kwon2024Fluctuationresponse,harunari2024Mutual,hasegawa2024Thermodynamic,aslyamov2024current,hasegawa2023Unifying,hasegawa2024Thermodynamic,kong2024fluctuation}.

Despite these developments, a comprehensive framework remains incomplete. Existing theories are largely confined to the linear-response regime where perturbations are assumed infinitesimal. This restriction is at odds with many realistic scenarios where systems routinely face finite parameter changes \cite{lan2012Energy, tu2018adaptation,Esposito2025rmp,mahdavi2024flexibility}. Another limitation is the predominant focus on trajectory-level observables \cite{dechant2020Fluctuation,kwon2024Fluctuationresponse,ptaszynski2024Dissipation,liu2025Dynamical,zheng2024Universal,aslyamov2024current,ptaszynski2024Nonequilibrium}. These results typically involve dynamical quantities such as activity, which are difficult to access experimentally. In contrast, more accessible state observables remain relatively underexplored, particularly in the nonlinear response regime, where a theoretical framework has been nonexistent.

\begin{figure}[!htb]
    \centering
    \includegraphics[width=1\columnwidth]{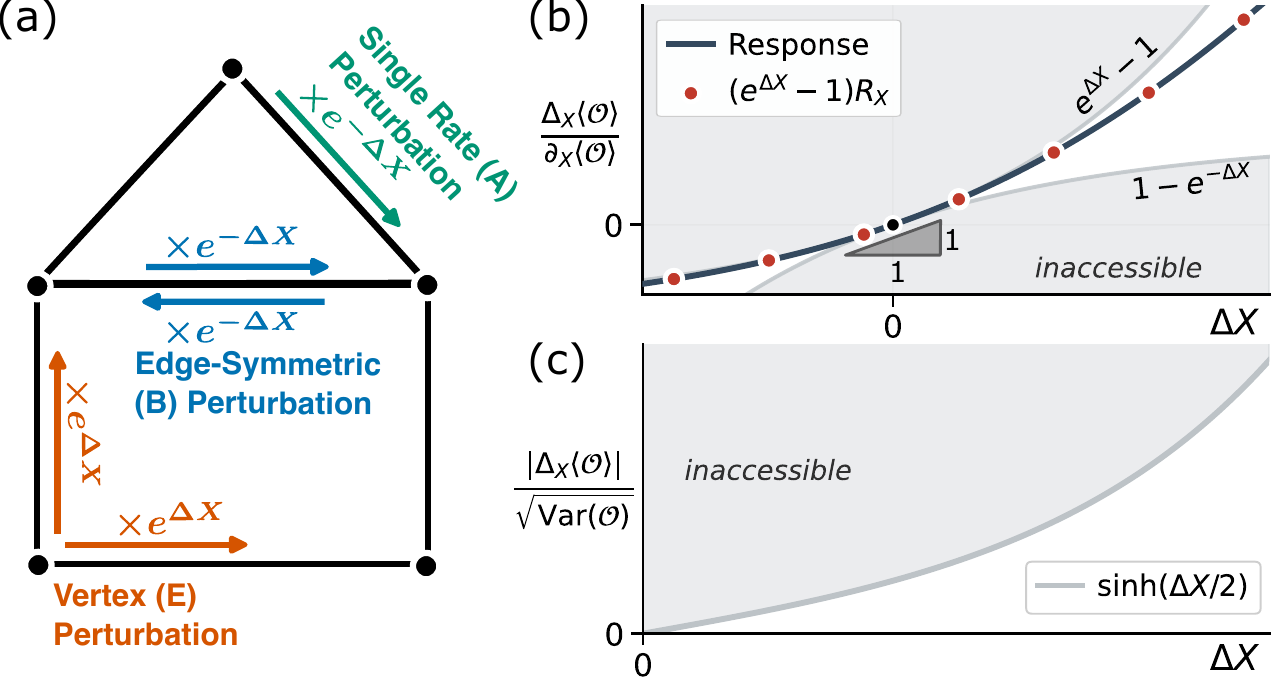}
    \caption{Exact identities and universal bounds for nonlinear response of steady-state observables. 
    (a) Schematic of three types of local perturbations to the transition rates of a continuous-time Markov chain.
    (b) The change of a state observable $\Delta_X\langle\mathcal{O}\rangle$ normalized by linear response $\partial_X\langle \mathcal{O}\rangle$ as a function of perturbation strength $\Delta X$. The exact response (black line) can be expressed with a scaling factor $R_X$ [Eq.~\eqref{eq:linear_nonlinear_identity}]. Universal bounds (gray lines) constrain all possible nonlinear responses [Eq.~\eqref{eq:bound_by_linear}].
    (c) Response curve normalized by the observable variance is upper bounded, revealing a universal bound on response precision [Eq.~\eqref{eq:response_resolution_bound}].}\label{fig:fig_main}
\end{figure}
{In this Letter, we address both of these challenges by developing a comprehensive theory for the response of continuous-time Markov chains under arbitrarily strong perturbations. We first introduce a general finite-perturbation identity that connects the steady-state probability distributions before and after a perturbation using the mean first-passage time (MFPT). We then parametrize the dynamics to focus on three archetypal local perturbations [Fig.~\ref{fig:fig_main} (a)] and derive an exact equality for any arbitrary observable. This equality reveals that the nonlinear response is directly proportional to the linear response via a scaling factor that depends on the change in local quantities. This relation allows us to establish a universal bound on the nonlinear response in terms of the linear response and perturbation strength [Fig.~\ref{fig:fig_main} (b)], as well as a bound on the relative response---termed the localization principle---which is saturated when the observable is localized. Finally, we establish an inequality relating the response and fluctuation of an arbitrary observable [Fig.~\ref{fig:fig_main} (c)], which shares the spirit of the fluctuation-dissipation theorem, providing a fundamental limit on response precision that holds for arbitrarily strong perturbations and systems arbitrarily far from equilibrium. We illustrate the applicability of our framework with a transcriptional control example. }

\textit{Setup and exact response relations for strong perturbations.}--- We consider a continuous-time Markov process with $N$ discrete states. The dynamics of the system is determined by a transition rate matrix $W$ whose off-diagonal element $W_{ij}$ denotes the transition rate on the edge $e_{ij}$ from state $j$ to state $i$. Its diagonal elements are defined as $W_{ii}=-\sum_{j(\neq i)}W_{ji}$. Assuming that $W$ is irreducible ensures the existence of a {unique} steady-state probability distribution $\boldsymbol{\pi}=(\pi_1,...,\pi_N)^{T}$ satisfying $W \cdot \boldsymbol{\pi}=\boldsymbol{0}$.

We now derive exact relations connecting steady-state responses to perturbations of arbitrary strength. Our starting point is a fundamental relation between perturbed and unperturbed steady-state probabilities:
\begin{equation}
\pi_k^{\prime}-\pi_k=\sum_{m}\sum_{n\neq m}\left(\langle t_{kn}\rangle - \langle t_{km}\rangle\right)\Delta W_{mn}\pi_n^{\prime}\pi_k,\label{Fmfpt}
\end{equation}
$\Delta W_{mn}=W_{mn}^{\prime}-W_{mn}$ is the change in transition rates, and {primed quantities (denoted by $'$) represent the corresponding quantities in the perturbed system}. This relation generalizes the response relations for infinitesimal perturbations developed in \cite{harvey2023Universal,aslyamov2024Nonequilibrium,ptaszynski2024Nonequilibrium} to finite-perturbation regime. Our derivation of Eq.~\eqref{Fmfpt} is solely based on the steady-state condition and the existence of a pseudoinverse of the generator, and is thus not limited to Markov jump processes, {see Supplemental Material (SM) \cite{supplemental_material} Sec. I for details} (see also \cite{khodabandehlou2024Affine} for a different approach and \cite{meyerjr.1980Condition,cho2000Markov} for discrete-time Markov chain). For $i\neq j$, $\langle t_{ij}\rangle$ represents the average time for the system to first reach state $i$ starting from state $j$, while we set $\langle t_{ii}\rangle=0$. 

To make this relation practical and connect it to physical perturbations, we parametrize the transition rates following \cite{dechant2020Fluctuation,owen2020Universal} as 
\begin{equation*}
    W_{ij}=e^{-A_{ij}},\quad A_{ij}=B_{ij}-E_{j}-F_{ij}/2.
\end{equation*} 
This decomposition captures the essential physics: { $A_{ij}$ represents the activation energy for transitions from state $j$ to state $i$}, decomposed into vertex parameters $E_j$ (energy well depths), symmetric edge parameters $B_{ij}=B_{ji}$ (barrier heights), and asymmetric edge parameters $F_{ij}=-F_{ji}$ (driving forces). Perturbations to $E_j$ affect all exit rates from state $j$, while edge perturbations modify specific transitions. The system is driven away from equilibrium whenever cycle affinities $F_c = \sum_{e_{ij}\in c}F_{ij}$ are non-zero. In this Letter, we do not impose any constraints on the cycle affinities; thus, the results hold for systems arbitrarily away from equilibrium.
 
{
Define the steady-state directed traffic $\phi_{mn}=W_{mn}\pi_n$
and current $j_{mn}=\phi_{mn}-\phi_{nm}$.
We perturb a \textit{single} parameter $X\to X'=X+\Delta X$, with arbitrary
$\Delta X\in\mathbb{R}$ and all other parameters fixed. Here, $X$ is either an edge parameter associated with a single chosen edge $e_{mn}$ or a vertex parameter associated with a single chosen vertex $m$. 
For $X=A_{mn}$ (A-type), $B_{mn}$ (B-type), or $E_m$ (E-type),
Eq.~\eqref{Fmfpt} gives, in each case, the exact steady-state
response $\pi_k\to\pi_k'$:
\begin{subequations}\label{eq:finite_main}
\begin{align}
   \text{A-type:\ }
    &\frac{\pi_k^{\prime}-\pi_k}{e^{\Delta A_{mn}}-1}=(\langle t_{km}\rangle-\langle t_{kn}\rangle )\pi_k \phi_{mn}',\label{finite0}\\
    \text{B-type:\ }&\frac{\pi_k^{\prime}-\pi_k}{e^{\Delta B_{mn}}-1}=(\langle t_{km}\rangle-\langle t_{kn}\rangle )\pi_k j_{mn}',\label{finite1}\\
    \text{E-type:\ }&\frac{\pi_k^{\prime}-\pi_k}{e^{\Delta E_m}-1}=\pi_m^{\prime}(\pi_k-\delta_{km})\label{finite3}
\end{align}
\end{subequations}
}where $\delta_{km}$ is the Kronecker delta. Eqs. \eqref{finite0}-\eqref{finite1} are new to our knowledge and Eq. \eqref{finite3} reproduces a key result in \cite{malloryigoshin2020,owen2020Universal}. The recursive relation $\sum_i W_{ij} \langle t_{ki}\rangle=\delta_{jk}/\pi_j-1$ and $\sum_iW_{ij}=0$ have been used to obtain \eqref{finite3}.
The generalization to multiple heat baths or multiple reaction channels can be found in { SM Sec. VIII} \cite{supplemental_material}. 

\begin{figure}[!thb]
     \centering
    \includegraphics[width=1\columnwidth]{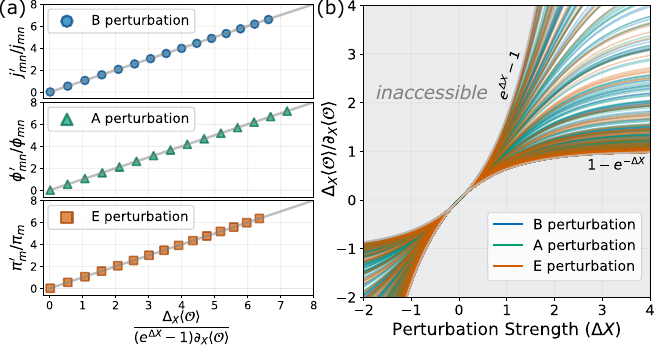}
    \caption{Response identities and bounds between linear and nonlinear responses. (a)~Numerical verification of the exact relations in Eq.~\eqref{eq:linear_nonlinear_identity}, which connect the nonlinear and linear responses of arbitrary observables to the change in local quantities. Scatter points correspond to different perturbation strengths. (b)~Numerical verification of the bound on nonlinear response from the linear response, Eq.~\eqref{eq:bound_by_linear}. Response curves are for a fully-connected 5-state network with randomly generated transition rates and observables.}
    \label{fig:nonlinear_linear_response}
\end{figure}

\textit{Nonlinear response identity.---
}Practical applications require understanding the response of experimentally
accessible observables \cite{fernandesmartins2023Topologically,chun2023Tradeoffs}. {For any state observable $\mathcal{O}$ with steady-state averages
$\langle \mathcal{O}\rangle=\sum_i \mathcal{O}_i\pi_i$
and
$\langle \mathcal{O}\rangle'=\sum_i \mathcal{O}_i\pi_i'$}, we present a novel framework for analyzing system responses to perturbations of arbitrary strength here.  From Eqs. \eqref{eq:finite_main}, we obtain exact relations between responses of state observables to strong (nonlinear) and weak (linear) perturbations as 
\begin{equation}\label{eq:linear_nonlinear_identity}
\begin{aligned}
    \frac{\Delta_{X}\langle \mathcal{O}\rangle}{e^{\Delta X}-1}=R_X\partial_X\langle \mathcal{O}\rangle, 
\end{aligned}
\end{equation}
where we define $\Delta_{X}\langle \mathcal{O}\rangle:=\langle \mathcal{O}\rangle'-\langle \mathcal{O}\rangle$ as the nonlinear response strength, {$\partial_{X}\langle \mathcal{O}\rangle:=\lim_{\Delta X\to 0}\Delta_X\langle\mathcal{O}\rangle/\Delta X$ is the linear response coefficient, and the scaling coefficients
\begin{equation}\label{eq:scaling_factor}
R_{A_{mn}}:=\frac{\phi_{mn}'}{\phi_{mn}},\quad R_{B_{mn}}:=\frac{j_{mn}'}{j_{mn}},\quad R_{E_{m}}:=\frac{\pi_m'}{\pi_m}.
\end{equation}
}
Equations~\eqref{eq:linear_nonlinear_identity}-\eqref{eq:scaling_factor} {constitute our first main result, the nonlinear response identity}, revealing that nonlinear responses to strong perturbations are proportional to linear responses, with the proportionality factor $R_X$ capturing the change in local quantities. This equality is numerically verified in Fig.~\ref{fig:nonlinear_linear_response}(a). {Notably, the scaling factors can be exactly expressed by the quantities of the unperturbed systems:
\begin{equation}\label{eq:perturbed_unperturbed}
    R_X=\frac{e^{-\Delta X}}{1+(e^{-\Delta X}-1)\alpha}.
\end{equation}
where the parameter $\alpha$ depends on the perturbation type: $\alpha = \phi_{mn}\langle t_{nm}\rangle$ for $X=A_{mn}$; $\alpha = \phi_{mn}\langle t_{nm}\rangle+\phi_{nm}\langle t_{mn}\rangle$ for $X=B_{mn}$; and $\alpha = \pi_m$ for $X=E_m$.}

Equation~\eqref{eq:linear_nonlinear_identity} has profound theoretical and practical implications. First, it demonstrates that nonlinear responses to arbitrarily strong perturbations can be exactly predicted using only the linear response coefficient and the perturbation-induced changes in local observables (flux, current, or probability). Second, Eqs.~\eqref{eq:linear_nonlinear_identity} are also of practical significance: change in local dynamical quantities (such as edge currents) can be inferred \textit{exactly} from coarse-grained global measurements of the response of \textit{any} global observable (a demonstrative application to a biochemical system is presented in Appendix~B).
{More strikingly, one can obtain the full response curve for any observable from simple measurements of the linear response $\partial_X\langle \mathcal{O}\rangle$ and $\langle \mathcal{O}\rangle'$ at an arbitrarily chosen $\Delta X$, without direct knowledge of dynamical details: Eq. \eqref{eq:linear_nonlinear_identity} can be rewritten as
\begin{equation*}
    \langle \mathcal{O}\rangle'=\frac{1-e^{-\Delta X}\partial_{X}\langle \mathcal{O}\rangle}{1+(e^{-\Delta X}-1)\alpha}+\langle \mathcal{O}\rangle.
\end{equation*}
This shows that such measurements determine the coefficient $\alpha$, which encodes the dynamical information of the unperturbed system. This is sufficient to fix the full response curve, i.e., to predict $\langle \mathcal{O}\rangle'$ for all $\Delta X$.}

\textit{Nonlinear response bound.
---}
The nonlinear response identity, Eq.~\eqref{eq:linear_nonlinear_identity}, links nonlinear and linear response via the scaling factor $R_X$. This factor can be determined exactly, either by measuring local quantities before and after a perturbation, or by predicting them from unperturbed system properties using Eq.~\eqref{eq:perturbed_unperturbed}. However, even without access to such detailed information, we now show that $R_X$ is universally bounded. This insight leads to a powerful result: the magnitude of the nonlinear response is itself bounded by the linear response.

To establish this, we first introduce a novel inequality between MFPTs and directed traffic ({see \cite{supplemental_material} Sec. III B for details}):
\begin{equation}\label{eq:flux-mfpt_bound}
   \langle t_{nm}\rangle+\langle t_{mn}\rangle\leq \frac{1}{\max\{ {\phi}_{mn},{\phi}_{nm}\}}.
\end{equation}
This inequality arises because two consecutive transitions from $m$ to $n$ must include at least one cycle through sequential first passages ($n\to m$ followed by $m\to n$) and thus take more time on average. 
Equality holds for two-state systems.  {More generally, Eq.~\eqref{eq:flux-mfpt_bound} is saturated whenever the transition $m\leftrightarrow n$ acts as an unavoidable kinetic gateway between the two endpoints; see \cite{supplemental_material} Sec.~III~B.} {Physically, the inequality implies that the more frequently transitions occur between two given states, the more kinetically indistinguishable they become.} Combining Eq.~\eqref{eq:flux-mfpt_bound} with the exact expressions in Eq.~\eqref{eq:perturbed_unperturbed} yields a universal bound on the scaling factor defined in Eq.~\eqref{eq:scaling_factor} ({see \cite{supplemental_material} Sec. IV for the proof}):
\begin{equation}\label{eq:scaling_factor_bound}
    \min\{e^{-\Delta X},1\}\leq R_X\leq \max\{e^{-\Delta X},1\},
\end{equation}
indicating that the change in local quantities is bounded by the perturbation strength. {In fact, $R_X$ is a monotone function of a single kinetic parameter $\alpha\in[0,1]$ as seen in Eq.~\eqref{eq:perturbed_unperturbed}, so the upper/lower bounds correspond to $\alpha\to 1$ and $\alpha\to 0$, respectively (\cite{supplemental_material} Sec.~IV).} 

Consequently, plugging Eq.~\eqref{eq:scaling_factor_bound} into the response identity Eq.~\eqref{eq:linear_nonlinear_identity},{ we establish the nonlinear response bound:}
\begin{equation}\label{eq:bound_by_linear}
    1-e^{-\Delta X}\leq\frac{\Delta_X\langle \mathcal{O}\rangle}{\partial_X\langle \mathcal{O}\rangle}\leq e^{\Delta X}-1.
\end{equation}
Numerically verifying these bounds reveals that the ratio of nonlinear to linear response is universally constrained by the perturbation strength alone, independent of system details. Fig.~\ref{fig:nonlinear_linear_response} (b) numerically verifies these universal bounds across different observables and perturbation strengths. 

\textit{Response localization principle.
}---
{Complete response characterization requires system details often unavailable in practice, making fundamental bounds essential for revealing universal properties with limited knowledge. Having bounded the nonlinear response using the linear response, we now seek more general bounds that hold even when the linear response is unknown. Since raw sensitivities can vary arbitrarily with units or scaling, we examine the relative response, which normalizes the response by the observable itself.}

We start from the driving law of localization, which emerges from the basic property of MFPTs. Note that Eq.~\eqref{eq:finite_main} relate the response to the difference in MFPTs between the two ends of the perturbed edge, and MFPTs always satisfy the triangle inequality ({ see \cite{supplemental_material} Sec. III C)}
\begin{equation}
    \langle t_{kn}\rangle \leq  \langle t_{km}\rangle+ \langle t_{mn}\rangle,
\end{equation}
{The equality holds only if all paths must pass through $m$, with any deviation reflecting alternative routes from $n$ to $k$ (see Sec.~II B in \cite{supplemental_material}).} We can thus bound the relative response of any state $k$ by the response at the two ends of the perturbed edge. Applying the inequality $\min_i(x_i/y_i)\leq \sum_ix_i/\sum_iy_i\leq \max_i(x_i/y_i)$ for $x_i\in \mathbb{R}$ and $y_i\geq 0$ (at least one positive $y_i$), we obtain the possible range of the relative response of an arbitrary observable in terms of the responses at the ends of the perturbed edge, with $\Delta X>0$ (if negative, we swap the upper and lower bounds):
\begin{equation}\label{eq:localization}
\frac{\pi_m^{\prime}-\pi_m}{\pi_m}\leq \frac{\Delta_{X}\langle \mathcal{O}\rangle}{\langle \mathcal{O}\rangle} \leq \frac{\pi_n^{\prime}-\pi_n}{\pi_n},
\end{equation}    
where $X=\{A_{mn},\,B_{mn}$\}. { See SM \cite{supplemental_material} Sec. V A for details.} For $B$-perturbation, we assume $j_{mn}'>0$ without loss of generality (if negative, we swap the indices $m$ and $n$). The upper and lower bounds always have opposite signs, as seen from { the main result \eqref{eq:linear_nonlinear_identity}}. 
This range reflects a fundamental localization principle: among all global observables, the relative response is maximized when the observable is localized at the endpoints of the perturbed edge. { The nonlinear response bound Eq.~\eqref{eq:bound_by_linear} together with the response localization principle Eq.~\eqref{eq:localization} is our second main result.}

This localization bound directly yields the range of relative response. Since the range equals the difference between responses at the two ends of the perturbed edge, we have
\begin{equation}\label{response_range}
    \begin{aligned}
        &\left|\frac{\pi_n^{\prime}}{\pi_n}-\frac{\pi_m^{\prime}}{\pi_m}\right|=R_X |e^{\Delta X}-1||\partial_X\ln(\pi_n/\pi_m)|,\\
        &\leq (e^{|\Delta X|}-1)|\partial_X\ln(\pi_n/\pi_m)|\leq e^{|\Delta X|}-1,
    \end{aligned}
\end{equation}
where the first equality provides an exact expression for the range of relative nonlinear response. The first inequality follows from the bound on the scaling factor [Eq.~\eqref{eq:scaling_factor_bound}], and the second inequality follows from the bound on linear response relation:
\begin{subequations}\label{eq:linear_relative_range}
    \begin{align}
        &|\partial_{B_{mn}}\ln(\pi_n/\pi_m)|= (\langle t_{mn}\rangle+\langle t_{nm}\rangle) |j_{mn}|\leq 1 \label{eq:linear_relative_range_B}\\
        &|\partial_{A_{mn}}\ln(\pi_n/\pi_m)|=(\langle t_{mn}\rangle+\langle t_{nm}\rangle)\phi_{mn}\leq 1 \label{eq:linear_relative_range_A}.
    \end{align}
\end{subequations}
where the upper bounds are obtained by applying Eq.~\eqref{eq:flux-mfpt_bound}. {This bound arises from the competition between two timescales: the local timescale in the perturbed edge, determined by the frequency of direct transitions, and the global timescale characterized by MFPTs.} For $X=B_{mn}$, Eq.~\eqref{eq:linear_relative_range_B} can be tightened by incorporating thermodynamics: using the constraint $|\ln({\phi}_{mn}/{\phi}_{nm})|\leq F^{\max}_{c_{mn}}$ \cite{liang2023Thermodynamic,liang2024ThermodynamicPRL}, we can tighten the upper bound by a factor of $(1-\exp(-F^{\max}_{c_{mn}}))$, where $F^{\max}_{c_{mn}}$ is the maximum cycle affinity over all cycles containing the perturbed edge; Using an enhanced inequality $|\partial_{B_{mn}}\ln (\pi_m/\pi_n)|\leq \tanh(F^{\max}_{c_{mn}}/4)$
derived in \cite{owen2020Universal} can be further tightened the bound by $\tanh(F^{\max}_{c_{mn}}/4)$.

Finally, since the upper and lower bounds of the relative response always have opposite signs in \eqref{response_range}, the absolute relative response is upper-bounded by $e^{|\Delta X|}-1$, yielding a universal bound $|\Delta_X\langle \mathcal{O}\rangle/\langle \mathcal{O}\rangle|\leq e^{|\Delta X|}-1$ for any observable under local perturbations. 

{\textit{Nonlinear response resolution limit.
}--- Beyond relative response, it is essential to quantify how reliably perturbations can be detected against fluctuations. This leads to the notion of response resolution, a signal-to-noise ratio that directly connects to the fluctuation-dissipation framework.} Here, $\langle \mathcal{O}\rangle:=\sum_i \mathcal{O}_i\pi_i$ is the steady-state average of an arbitrary observable $\mathcal{O}$ and $\text{Var}(\mathcal{O}):=\sum_i(\mathcal{O}_i-\langle \mathcal{O}\rangle)^2\pi_i$ is its steady-state variance. 

We establish a universal bound for finite perturbations by directly applying the Cauchy-Schwarz inequality to the response expression. Decomposing the finite response as $\Delta_X\langle \mathcal{O}\rangle = \sum_i(\mathcal{O}_i-\langle\mathcal{O}\rangle)(\pi_i'-\pi_i)$ and introducing the probability weights $\sqrt{\pi_i}$, we obtain: $\Delta_X\langle \mathcal{O}\rangle = \sum_i\left[(\mathcal{O}_i-\langle\mathcal{O}\rangle)\sqrt{\pi_i}\right]\left[\sqrt{\pi_i}(\pi_i'/\pi_i-1)\right] 
\leq \sqrt{\text{Var}(\mathcal{O})}\sqrt{\sum_i \pi_i(\pi_i'/\pi_i-1)^2}$, where the second factor is the $\chi^2$-divergence between the perturbed and unperturbed probability distributions, i.e. the $\chi^2(\pi'||\pi)=\langle \epsilon^2\rangle$ with $\epsilon_i=\pi_i'/\pi-1$ defined as the relative response. The key insight is that for edge perturbations ($X \in \{A_{mn}, B_{mn}\}$), the relative probability changes $\epsilon_i = \pi_i'/\pi_i - 1$ satisfy universal constraints. Since $\langle \epsilon\rangle = 0$ and the response is localized at the edge endpoints [Eq.\eqref{eq:localization}], we can get universal bound on the $\chi^2$-divergence: $\sqrt{\langle\epsilon\rangle^2}=\sqrt{\text{Var}(\epsilon)} \leq \sqrt{|\epsilon_m||\epsilon_n|} \leq |\sinh(\Delta X/2)|=|(e^{\Delta X/2}-e^{-\Delta X/2})/2|$, where the first inequality follows the bound on variance of bounded variables \cite{bhatia2000Better}, and the final inequality follows from MFPT inequalities and the explicit expressions for $\epsilon_m$ and $\epsilon_n$ {(\cite{supplemental_material} Sec. VI)}. For vertex perturbations ($X = E_m$), {a direct calculation using Eq. \eqref{Fmfpt} yields the same bound (see \cite{supplemental_material} Sec. VI C)}. This yields our {third main result}, the universal bound on response resolution:
\begin{equation}\label{eq:response_resolution_bound}
    \frac{|\Delta_X\langle \mathcal{O}\rangle|}{\sqrt{\text{Var}(\mathcal{O})}}\leq \left|\sinh\left(\frac{\Delta X}{2}\right)\right|.
\end{equation}
{The bound exhibits a fundamental symmetry: since we can equivalently view the ``original'' system as the ``perturbed'' one with perturbation $-\Delta X$, the normalization can use either $\sqrt{\text{Var}(\mathcal{O})}$ or $\sqrt{\text{Var}'(\mathcal{O})}$. The bound depends solely on perturbation strength $|\Delta X|$ and is independent of system size, network topology, or observable choice. For weak perturbations ($|\Delta X| \ll 1$), it reduces to $|\Delta X|/2$ and leads to the linear response precision bound $|\partial_X\langle \mathcal{O}\rangle|/\sqrt{\text{Var}(\mathcal{O})}\leq 1/2$. {Notably, the linear-response bound can also be derived from the results in \cite{fernandesmartins2023Topologically}.} For strong perturbations it grows as $e^{|\Delta X|/2}/2$, establishing fundamental limits across all perturbation regimes. This reveals the physical significance of Eq.~\eqref{eq:response_resolution_bound} as a \textit{response resolution limit}: it gives a universal upper bound for distinguishing observable values before and after perturbation against intrinsic fluctuations, establishing the ultimate signal-to-noise ratio with which finite perturbations can be detected in Markovian systems. {Eq.~\eqref{eq:response_resolution_bound} is mathematically sharp. It saturates in effective two-state regimes where the perturbed element is a kinetic gateway and observables align with the relative steady-state change $\epsilon_i=\pi_i'/\pi_i-1$ (see \cite{supplemental_material} Sec.~VI~D). Therefore, slack indicates network redundancy or sub-optimal observables.} Incorporating thermodynamics can further improve the bound for B-perturbation by a factor 
$\tanh(F^{\max}_{c_{mn}}/4)$ {(\cite{supplemental_material} Sec. VI B)}. Multi-edge generalization of Eq. \eqref{eq:response_resolution_bound} is presented in Appendix A.}

\begin{figure}[!t]
    \centering
    \includegraphics[width=1\columnwidth]{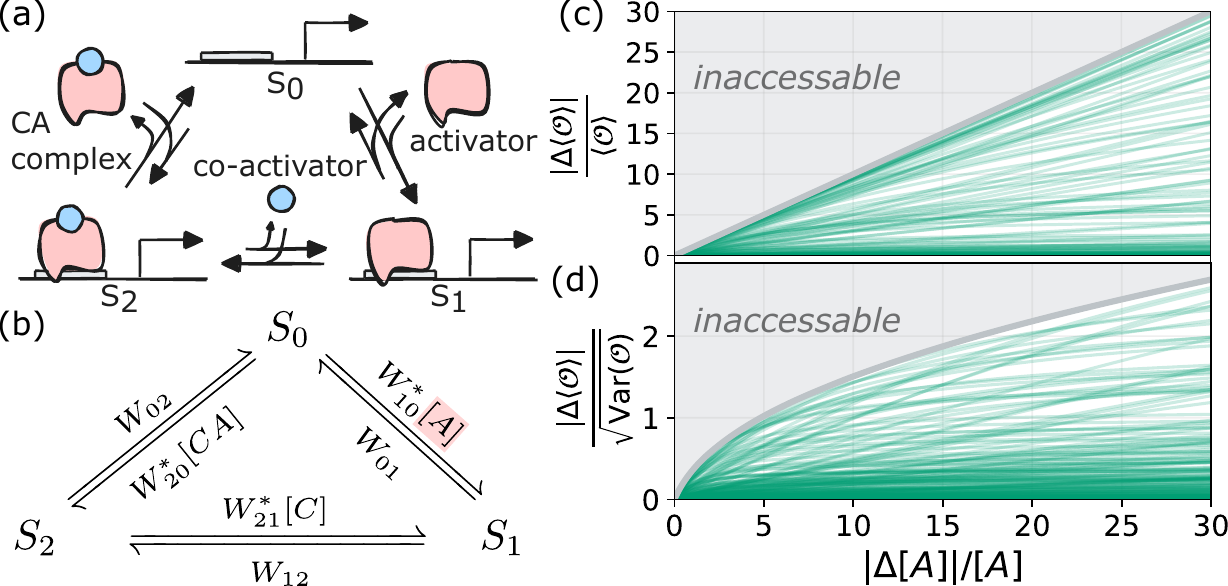}
    \caption{Transcriptional regulation model and its response bounds. (a) The three-state promoter model controlled by activator A and co-activator C. (b) The Markov chain representation of the model, where the activator concentration controls the transition rate from state 0 to state 1. (c) The relative response of expression level is upper bounded by the relative change of activator concentration. (d) The response resolution bound for the expression level is universally bounded. For (c) and (d), we generate three-state networks with random kinetics and randomly assign expression levels to each state, and perturb the activator concentration to get the response curves.}
    \label{fig:regulation}
\end{figure}

\textit{Application: Transcriptional Regulation.}--- Biological information processing operates under fundamental physical constraints \cite{tenwolde2016Fundamental}. We demonstrate this using a paradigmatic gene regulation process \cite{phillips2020Molecular} where activator A and co-activator C control transcription through a three-state promoter: $S_0$ (basal), $S_1$ (A-bound), and $S_2$ (AC-bound), as shown in Fig.~\ref{fig:regulation}(a). The expression level $\langle \mathcal{O}\rangle=\sum_{i=0}^2\mathcal{O}_i\pi_i$ depends on state-specific rates $\mathcal{O}_i$. Modulating activator concentration $[A]$ perturbs the binding rate $W_{10}=W_{10}^* [A]$ [Fig.~\ref{fig:regulation}(b)], yielding a single-edge perturbation with $\Delta X = -\Delta\ln[A]=\ln([A]/[A]')$. Our framework provides two complementary bounds. First, the relative response bound yields $|\Delta\langle\mathcal{O}\rangle/\langle\mathcal{O}\rangle|\leq e^{|\Delta X|}-1=|\Delta[A]/[A]|$, establishing that the relative change in expression level never exceeds the relative change in activator concentration. Second, the response resolution bound [Eq.~\eqref{eq:response_resolution_bound}] yields $\frac{|\Delta\langle \mathcal{O}\rangle|}{\sqrt{\text{Var}(\mathcal{O})}}\leq \left|\sqrt{[A]'/[A]}-\sqrt{[A]/[A]'}\right|/2$. For expression changes to reliably exceed intrinsic noise, ${|\Delta\langle \mathcal{O}\rangle|} > {\sqrt{\text{Var}(\mathcal{O})}}$, we require $|\Delta [A]/[A]| \gtrsim4.8$. {The bound is independent of detailed kinetics and becomes tight when the perturbed transition acts as a kinetic gateway and the observable is aligned with the perturbation (see the saturation condition discussion for Eq.~\ref{eq:response_resolution_bound} in SM Sec.~VI~D \cite{supplemental_material}). Any violation of the bound falsifies single transition control, implying that the input effectively modulates multiple transitions and thereby enabling inference on the number of edges influenced.}

\textit{Discussion.}---We have established a comprehensive theory for nonequilibrium nonlinear response, grounded in a novel connection between steady-state perturbations and mean first-passage times. Our framework provides a hierarchy of exact relations and universal bounds applicable to arbitrarily strong perturbations, with predictive power scaling with the level of available information.
{The derivations exploit physical constraints of MFPTs, lending the resulting bounds a direct kinetic interpretation and transparent saturation conditions tied to the gateway structure of the network.}

{Our results form a predictive hierarchy. At the most fundamental level, full knowledge of the unperturbed system (including MFPTs) allows for the exact prediction of the response to any finite perturbation [Eqs.~\eqref{eq:linear_nonlinear_identity}-\eqref{eq:perturbed_unperturbed}]. More practically, our cornerstone identity [Eq.~\eqref{eq:linear_nonlinear_identity}] links nonlinear to linear response via a simple scaling factor, enabling a powerful two-way inference between local dynamics and global observations {(as demonstrated in Appendix B for inferring ATP consumption rates)}, and offering a new perspective on inference in stochastic systems \cite{seifert2019Stochastic,HDPR22PRX,MES22PRX,15PRL_TUR,16PRL_TUR,horowitz2020Thermodynamic}. The identity further leads to a cascade of universal bounds requiring progressively less information: the nonlinear response is bounded by the linear [Eq.~\eqref{eq:bound_by_linear}], while the relative response [Eq.~\eqref{response_range}] and the signal-to-noise ratio are bounded even without linear response information [Eq.~\eqref{eq:response_resolution_bound}], the latter serving as a strong-perturbation analogue of the fluctuation-dissipation inequality and imposing a fundamental limit on response resolution.}

Our framework enhances and extends recent bounds on nonequilibrium response in a unified way \cite{owen2020Universal,owen2023Size,fernandesmartins2023Topologically,aslyamov2024Nonequilibrium}. Notably, the framework is not limited to state observables. {The main identity \eqref{eq:linear_nonlinear_identity} also holds for any current observable (see Appendix C). }
Furthermore, our exact identities provide a platform to generalize other key results to the strong-perturbation regime, such as the fluctuation-response relations in Ref.~\cite{aslyamov2024current,ptaszynski2024Nonequilibrium} (see Appendix D).

Looking forward, it would be valuable to extend our theory to systems governed by Lindblad master equations. Furthermore, addressing different classes of perturbations, such as time-dependent signals commonly encountered in biological systems \cite{nicoletti2024information,nicoletti2024Informationelife}, represents a promising avenue for future research. Investigating the response of transient dynamics, as studied in \cite{mitrophanov2003stability,mitrophanov2004spectral,mitrophanov2024}, is another intriguing direction. Finally, our results are generalizable to continuous-space systems governed by the Fokker-Planck equation, using the methods in \cite{gao2022thermodynamic,sawchuk2024dynamical}.

\begin{acknowledgments}  
R.B. was supported by JSPS KAKENHI Grant No. 25KJ0766. We are grateful to the Statistical Physics Youth Communication community for providing the initial platform for this discussion. We thank Timur Aslyamov, Marco Baiesi, Christian Maes, and Jiming Zheng for their helpful comments on the manuscript. R.B. thanks Faezeh Khodabandehlou for communications. We thank Kangcan Yao for pointing out some typos. We also acknowledge the assistance of Claude and ChatGPT for their help with writing, critical comments, and inspiring discussions.
\end{acknowledgments} 

\appendix

\section*{End Matter}

\textit{Appendix A: Nonlinear response to multi-edge perturbations.---}In the more general setting of simultaneous perturbations to $m$ different edges, we can sequentially analyze the effect by adding perturbations one by one. Let $\langle \mathcal{O}\rangle^{(k)}$ denote the updated mean value after the $k$-th edge perturbation. We notice that the nonlinear response to perturbation on $m$ different edges can be expressed as 
\begin{align}\label{multi_equality}
    \langle \mathcal{O}\rangle^{\prime}-\langle\mathcal{O}\rangle&=\sum_{k=1}^{m}(\Delta_{X_{k-1}}\langle \mathcal{O}\rangle)\nonumber\\
    &=\sum_{k=1}^m(e^{\Delta_{X_{k-1}}}-1)R_{X_{k-1}}\partial_{X_{k-1}} \langle \mathcal{O}\rangle,
\end{align}
where $\langle \mathcal{O}\rangle^{\prime}\equiv \langle \mathcal{O}\rangle^{(m)},\ \langle \mathcal{O}\rangle^{(0)}\equiv\langle\mathcal{O}\rangle$, $\Delta_{X_{k-1}}\langle \mathcal{O}\rangle\equiv \langle \mathcal{O}\rangle^{(k)}-\langle \mathcal{O}\rangle^{(k-1)}$ and $R_{X_{k-1}}$ has the same definition as in Eq. \eqref{eq:scaling_factor}. Eq. \eqref{multi_equality} generalizes Eq. \eqref{eq:linear_nonlinear_identity} to multi-edge perturbations. 

Given that $\Delta_{X_{k-1}}\langle \mathcal{O}\rangle$ is nothing but the nonlinear response to the single-edge perturbation, the bounds for nonlinear response to multi-edge perturbation are readily obtained by combining the single-edge bounds in the main text with the triangle inequality. For example, Eqs. \eqref{eq:bound_by_linear} and \eqref{eq:response_resolution_bound} are generalized to
\begin{align}
&|\langle \mathcal{O}\rangle'-\langle \mathcal{O}\rangle|\nonumber\\
&\leq m \max_k|\partial_{X_k}\langle \mathcal{O}\rangle||e^{\Delta X_{\rm max}}-1|
\leq m|e^{\Delta X_{\rm max}}-1|, \\
    &\frac{|\langle \mathcal{O}\rangle^{\prime}-\langle \mathcal{O}\rangle|}{\sqrt{\text{Var}^{\rm max}(\mathcal{O})}}\leq m\left|\sinh\left(\frac{\Delta X_{\rm max}}{2}\right)\right|,
\end{align}
where $|\Delta X_{\rm max}|$ is the maximum perturbation strength and $\text{Var}^{\rm max}(\mathcal{O}):=\max_k\text{Var}^{(k)}(\mathcal{O})$ is the maximum variance. This may be interpreted as an upper bound for the worst signal-to-noise ratio. In the linear response limit, the upper bound will be proportional to the number of perturbed edges, as known.

\textit{Appendix B: Application to biochemical sensing}---
Our nonlinear response identity, Eq.~\eqref{eq:linear_nonlinear_identity}, provides a powerful link between nonlinear response and local system properties. Here, we demonstrate its practical utility in the context of biochemical sensing, showing how it enables a two-way inference between global observables and local dynamics.

\begin{figure}[!htb]
    \centering
    \includegraphics[width=1\columnwidth]{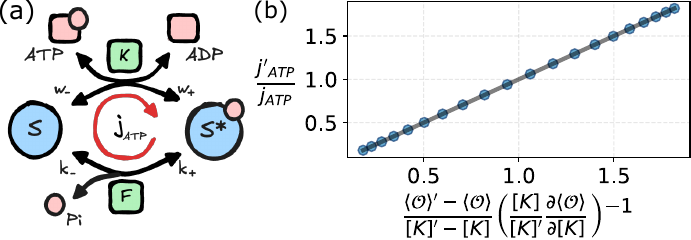}
    \caption{{Inferring metabolic change from global response in a push-pull motif.} (a) A canonical push-pull signaling circuit powered by ATP consumption. (b) The exact relation between the response of any global observable $\langle\mathcal{O}\rangle$ and the change in the local ATP consumption current. This allows the fractional change in metabolic cost ($j_{\text{ATP}}'/j_{\text{ATP}}$) to be precisely inferred from measurements of the system's global response to a finite change in kinase concentration. Here we take $w_+=k_-=1$ and $w_-=k_+=e^{-1}$, and $[K]=[F]$, while varying $[K]'/[K]$ between 0.1 and 10. The observable is defined as $\mathcal{O}_S=-1$ and $\mathcal{O}_{S^*}=2$.}
    \label{fig:push_pull}
\end{figure}

We consider a canonical push-pull network, a ubiquitous motif in cellular signaling \cite{li2003Sensitivity,bialek2005Physical,owen2020Universal,govern2014Energy}. In this model, a substrate interconverts between an active ($S^*$) and inactive ($S$) state, catalyzed by a kinase ($K$) and a phosphatase ($F$), respectively [Fig.~\ref{fig:push_pull}(a)]. To connect this system to our framework, we note that the kinase-catalyzed rate depends on the kinase concentration $[K]$, corresponding to an effective energy barrier $B_w \propto -\ln[K]$. A finite change in $[K]$ thus causes an edge-symmetric B-perturbation with $\Delta B_w = \ln([K]/[K]')$. The current along this edge, $j_w$, is precisely the net rate of ATP consumption, $j_{\text{ATP}}$.

Substituting these relations into our main identity, Eq.~\eqref{eq:linear_nonlinear_identity}, yields a direct connection to the ATP consumption rate. This holds for any system observable $\langle \mathcal{O} \rangle$, which serves as a readout of the signaling process -- typically a linear combination of the active and inactive substrate concentrations. This leads to the exact relation:
\begin{equation}
   \frac{\Delta_{[K]}\langle \mathcal{O}\rangle}{[K]/[K]'-1} = -\frac{j_{\text{ATP}}'}{j_{\text{ATP}}}\frac{\partial\langle \mathcal{O}\rangle}{\partial\ln[K]}.
\end{equation}
This identity allows for a powerful two-way inference, as illustrated in Fig.~\ref{fig:push_pull}(b). An experimentalist can measure the response of an accessible global readout to a known change in kinase concentration to precisely \textit{infer} the change in the system's metabolic cost ($j_{\text{ATP}}'/j_{\text{ATP}}$). Conversely, this relation can be used to \textit{predict} the response of any observable if the change in metabolic cost is known.

\textit{Appendix C: Nonlinear response relations for current observables.---}We now extend { the main identity \eqref{eq:linear_nonlinear_identity}} to current observables. {Consider a general current observable $\langle \mathcal{J}\rangle:=\sum_{\substack{m, n \\ m < n}} \mathcal{J}_{mn}j_{mn}$. For the same set of perturbation on $X=B_{mn},\ A_{mn},\ E_m$ ($X\to X':=X+\Delta X$), an arbitrary current observable satisfies the same nonlinear response relation (see \cite{supplemental_material} Sec. X), namely:
\begin{equation}
    \frac{\langle \mathcal{J}\rangle^{X'}-\langle \mathcal{J}\rangle^{X}}{e^{\Delta X}-1}= R_{X} \frac{\partial\langle \mathcal{J}\rangle}{\partial X}.
\end{equation}
This exact relation can be utilized to generalize some previously obtained results for linear response to nonlinear response scenarios. For instance, the fluctuation-response relations and the response thermodynamic uncertainty relation developed in \cite{aslyamov2024current,ptaszynski2024Dissipation}.
 }
 
We then derive exact expressions for steady-state current responses and their upper bounds for three types of perturbation:
\begin{subequations}
\begin{align}
     &\frac{\partial j_{kl}}{\partial B_{mn}}=-[\delta_{(k,l)}^{(m,n)}+\Delta t_{nm}^l{\phi}_{kl}-\Delta t_{nm}^k{\phi}_{lk}]j_{mn},\label{CRR1}\\
     & \frac{\partial j_{kl}}{\partial A_{mn}}=[\delta_{(k,l)}^{(m,n)}+\Delta t_{nm}^l{\phi}_{kl}-\Delta t_{nm}^k{\phi}_{lk}]{\phi}_{mn},\label{CRRF1} \\
     & \frac{\partial j_{kl}}{\partial E_{m}}=\pi_m j_{kl},
\end{align}
\end{subequations}
where $\Delta t_{nm}^l:=\langle t_{ln}\rangle-\langle t_{lm}\rangle$, and $ \delta_{(k,l)}^{(m,n)} = \delta_{mk}\delta_{nl} -\delta_{ml}\delta_{nk}$ is edge delta function to distinguish local and non-local perturbations. 

Then, using Eq.~\eqref{eq:flux-mfpt_bound}, we obtain upper bounds for non-local and local current response as (take $B_{kl}$ as an example) 
\begin{subequations}

\begin{align}
    &\left|\frac{\partial j_{kl}}{\partial B_{mn}} \right|\leq 
    \frac{|j_{mn}|a_{kl}}{\max(\phi_{mn},\phi_{nm})},\label{currentbound1}\\
    &0\leq \frac{\partial j_{mn}}{\partial B_{mn}}\frac{\left \vert j_{mn}\right \vert}{j_{mn}} \leq  |j_{mn}|\leq {a}_{mn},\label{currentbound2}
\end{align}
\end{subequations}
{where $a_{mn}=\phi_{mn}+\phi_{nm}$ is the activity on edge $mn$.} The non-local perturbation bound is new to our knowledge, and the local perturbation bound recovers the result in \cite{aslyamov2024Nonequilibrium}.

Thus, the single-edge response for an arbitrary current observable $\langle \mathcal{J}\rangle:=\sum_{k<l}\mathcal{J}_{kl}j_{kl}$
is upper-bounded as
\begin{equation}
    \frac{\partial\langle \mathcal{J}\rangle}{\partial B_{mn}}\leq \sum_{k<l}|\mathcal{J}_{kl}|{a}_{kl}.\label{currentbound3}
\end{equation}
Unlike steady-state distributions and state observables, current response is more intrinsically linked to activity. Eqs.~\eqref{currentbound1}-\eqref{currentbound3} show that higher activity leads to enhanced responses in both currents and current observables. This occurs because current observables inherently depend on jump frequencies, while state observables only depend on residence times. However, if normalizing the current observable using the escape rates, their response precision can be upper bounded by a size-independent constant, $3/2$, similar to the state observables (\cite{supplemental_material} {Sec. IX}).

\textit{Appendix D: Fluctuation-response relations for arbitrarily strong perturbations.---}We now show another corollary of our central results. Using Eqs. \eqref{eq:linear_nonlinear_identity}, we extend the fluctuation-response relation for {current} and state observables 
derived in \cite{aslyamov2024current,ptaszynski2024Nonequilibrium} to the finite-perturbation regime:
\begin{equation}
\begin{aligned}
    &\text{Cov}[\mathcal{O}^{(1)},\mathcal{O}^{(2)}]\\&=\sum_{m<n} \frac{{a}_{mn}}{(j^{B_{mn}^{\prime}}_{mn})^2}\left(\frac{\Delta_{B_{mn}} \langle \mathcal{O}^{(1)}\rangle}{e^{\Delta B_{mn}}-1}\right)\left(\frac{\Delta_{B_{mn}} \langle \mathcal{O}^{(2)}\rangle}{e^{\Delta B_{mn}}-1}\right)\\
    &=\sum_{m,n}\frac{\phi_{mn}}{(\phi_{mn}^{\prime})^2}\left(\frac{\Delta_{A_{mn}} \langle \mathcal{O}^{(1)}\rangle}{e^{\Delta A_{mn}}-1}\right)\left(\frac{\Delta_{A_{mn}} \langle \mathcal{O}^{(2)}\rangle}{e^{\Delta A_{mn}}-1}\right),
\label{FFRR1}
\end{aligned}
\end{equation}
{where $\mathcal{O}^{(1)},\ \mathcal{O}^{(2)}$ are of the same type, being either current or state observables.}

When the perturbations are small, relations in \cite{aslyamov2024current,ptaszynski2024Nonequilibrium} are recovered. When $\mathcal{O}^{(1)}=\mathcal{O}^{(2)}$, all terms on the right-hand side of Eq. \eqref{FFRR1} are non-negative, so they provide natural lower bounds for the covariance, establishing strong-perturbation variants of fluctuation-response inequalities \cite{ptaszynski2024Nonequilibrium,kwon2024Fluctuationresponse}. 

\bibliography{refs}
\end{document}


\begin{CJK*}{UTF8}{gbsn}
\title{Supplementary Material for ``Nonlinear Response Identities and Bounds for Nonequilibrium Steady States''
}

\author{Ruicheng Bao}
\email{Corresponding author: ruicheng@g.ecc.u-tokyo.ac.jp}
\affiliation{Department of Physics, Graduate School of Science, 7-3-1, 
The University of Tokyo, Hongo, Bunkyo-ku, Tokyo 113-0033, Japan}

\author{Shiling Liang (梁师翎)}
\email{Corresponding author: shiling@pks.mpg.de}
\affiliation{Center for Systems Biology Dresden, 01307 Dresden, Germany}
\affiliation{Max Planck Institute for the Physics of Complex Systems, 01187 Dresden, Germany}
\affiliation{Max Planck Institute of Molecular Cell Biology and Genetics, 01307 Dresden, Germany}
\maketitle
\end{CJK*}


\onecolumngrid
This supplementary material contains the detailed derivation and proof of the results in the main text.
\vspace{12pt}

\tableofcontents
\section{Derivation of the general response equality [Eq.~(1) in main text] using Drazin pseudoinverse\label{sec:response_equality}}
We first provide the definitions of MFPT and Drazin pseudoinverse of rate matrix, and an explicit expression of the MFPT using Drazin pseudoinverse of the rate matrix.
The MFPT from state $j$ to state $i$ is defined as  
\begin{equation}
    \langle t_{ij}\rangle = \int_0^{\infty} t_{ij} f(t_{ij})dt_{ij},
\end{equation}
where $t_{ij}$ is the random first passage time and $f(t_{ij})$ is its probability density function. $t_{ij}$ is a random variable defined as 
\begin{equation*}
    t_{ij} := \inf_{t\geq 0}\{X(t) = i | X(0) = j\}.
\end{equation*}
Here, $X(t)$ is the state of the system at time $t$. That is, $t_{ij}$ is the first time at which a stochastic trajectory reaches state $i$ at time $t$, given that it starts from state $j$ at time 0. The MFPT has a closed-form expression using the Drazin pseudoinverse \cite{harvey2023Universal,yao1985first},
\begin{equation}
     \langle t_{ij}\rangle=\frac{W_{ij}^{\mathcal{D}}-W^{\mathcal{D}}_{ii}}{\pi_i},\label{drazinmfpt}
\end{equation}
which will be used later. The Drazin pseudoinverse $W^{\mathcal{D}}$ of a rate matrix $W$ is defined as the matrix satisfying the following conditions \cite{campbell2009generalized}:
\begin{align*}
    & W^{\mathcal{D}}WW^{\mathcal{D}}=W^{\mathcal{D}},\\
    &W^{\mathcal{D}}W=WW^{\mathcal{D}},\\
    &WW^{\mathcal{D}}W=W.
\end{align*}
A useful property of $W^{\mathcal{D}}$ is \cite{campbell2009generalized}
\begin{equation}
    W^{\mathcal{D}}W=\mathbb{I}-\boldsymbol{\pi}\boldsymbol{e},
\end{equation}
where $\mathbb{I}$ is an $N\times N$ identity matrix, $\boldsymbol{\pi}$ is written as a column vector and $\boldsymbol{e}=(1,...,1)$ is a row vector with all $N$ elements being 1. 
A possible choice of $W^{\mathcal{D}}$ is
\begin{equation*}
    W^{\mathcal{D}}=-\int_0^{\infty}(e^{Wt}-\boldsymbol{\pi}\boldsymbol{e})dt.
\end{equation*}

In what follows, we derive Eq. (1) in the main text. We note that a specific case (only for linear response to infinitesimal perturbation) of Eq. (1) was first derived in \cite{harvey2023Universal}, and an equivalent form of it was recently re-derived in \cite{ptaszynski2024Critical,ptaszynski2024Nonequilibrium}, without identifying the MFPT from Drazin pseudoinverse. 

We denote the transition rate matrices before and after an arbitrarily strong perturbation as $W$ and $W^{\prime}$. Similarly, column vectors $\boldsymbol{\pi}$ and $\boldsymbol{\pi}^{\prime}$ are the steady-state probability distribution before and after the perturbation. Observe that
\begin{equation}
    W(\boldsymbol{\pi}^{\prime}-\boldsymbol{\pi})=-(W^{\prime}-W)\boldsymbol{\pi}^{\prime} \label{S1}
\end{equation}
because of the steady-state condition $W\boldsymbol{\pi}=\boldsymbol{0}$ and $W^{\prime}\boldsymbol{\pi}^{\prime}=\boldsymbol{0}$. We aim to express the finite response $\boldsymbol{\pi}^{\prime}-\boldsymbol{\pi}$ using the perturbation $W^{\prime}-W$ on transition rate matrix, so that the $W$ on the left should be put to the right. Since $W$ is not invertible due to its one-dimensional null space ($W\boldsymbol{\pi}=\boldsymbol{0}$ and $W\boldsymbol{e}^T=0$), we should resort to the Drazin  $W^{\mathcal{D}}$. Multiplying Eq.~\eqref{S1} by $W^{\mathcal{D}}$ on both sides, we obtain
\begin{equation}
    \boldsymbol{\pi}^{\prime}-\boldsymbol{\pi}=-W^{\mathcal{D}}(W^{\prime}-W)\boldsymbol{\pi}^{\prime}, 
\end{equation}
where the property of Drazin pseudoinverse, $W^{\mathcal{D}}W=WW^{\mathcal{D}}=\mathbb{I}-\boldsymbol{\pi}\boldsymbol{e}$ has been used. We also use $\boldsymbol{e}(\boldsymbol{\pi}^{\prime}-\boldsymbol{\pi})=0$ from normalization. Consider a minimal case when a single transition rate $W_{ij}$ is perturbed to be $W^{\prime}_{ij}$. In this case, the matrix $W^{\prime}-W$ only has two non-zero entries $(W^{\prime}-W)_{ij}=\Delta W_{ij}$ and $(W^{\prime}-W)_{jj}=-\Delta W_{ij}$, which are both in the $j$-th column. Then, the $k$-th row of the matrix $-W^{\mathcal{D}}(W^{\prime}-W)$ would be

\begin{equation}  
[-W^{\mathcal{D}}(W^{\prime}-W)]_k=(0 \; \cdots \; \stackrel{\textcolor{blue!70!black}{j-th \ \text{element}}}{(W^{\mathcal{D}}_{kj} -W^{\mathcal{D}}_{ki} )\Delta W_{ij}}  \; \cdots \; 0), 
\end{equation}
which has a single non-zero element, the $j$-th one. Consequently, the $k$-th element of the column vector $\boldsymbol{\pi}^{\prime}-\boldsymbol{\pi}$ is 
\begin{align}
    \pi^{\prime}_k-\pi_k&=(W^{\mathcal{D}}_{kj} -W^{\mathcal{D}}_{ki} )\Delta W_{ij}\pi^{\prime}_j\nonumber \\
    &=(\langle t_{kj}\rangle - \langle t_{ki}\rangle)\Delta W_{ij}\pi_j^{\prime}\pi_k,
\end{align}
where Eq.~\eqref{drazinmfpt} has been used. For general cases, $[-W^{\mathcal{D}}(W^{\prime}-W)]_k$ becomes
\begin{equation}
    (\sum_{i=1}^N\overset{\textcolor{blue!70!black}{1-st}}{(W^{\mathcal{D}}_{k1} -W^{\mathcal{D}}_{ki} )\Delta W_{i1}}\; \cdots \sum_{i=1}^N\overset{\textcolor{blue!70!black}{j-th}}{(W^{\mathcal{D}}_{kj} -W^{\mathcal{D}}_{ki} )\Delta W_{ij}} \cdots).
\end{equation}
That is, adding contributions from single transition rate perturbations up gives rise to the desired result
\begin{equation}
    \pi_k^{\prime}-\pi_k=\sum_{\substack{m, n \\ m < n}} (\langle t_{kn}\rangle - \langle t_{km}\rangle)(\Delta W_{mn}\pi_n^{\prime}-\Delta W_{nm}\pi_m^{\prime})\pi_k.\label{eqs:Fmfpt}
\end{equation}
which is the Eq.~(1) in the main text. A discrete-time version of Eq.~\eqref{eqs:Fmfpt} was derived in \cite{meyerjr.1980Condition}. We remark that the discrete-time result cannot be transformed into Eq.~\eqref{eqs:Fmfpt} by normalizing transition rates of the continuous-time process.

The derivation provided here is not limited to Markov jump processes. Generalization to the Fokker-Planck equation is possible, given that one can define a pseudoinverse of the Fokker-Planck operator \cite{sawchuk2024dynamical}.

\section{Derivation of exact response relation for parameterized perturbations [Eq.~(3) in the main text]}
In the main text, we introduced the parameterization of transition rates 
\begin{equation}\label{eqs:rate_parameterizaiton}
W_{ij} = e^{-A_{ij}} \text{ with } A_{ij} =B_{ij} -E_j-F_{ij}/2,
\end{equation}
and considering three types of perturbations, single rate perturbation $A_{mn}'=A_{mn}+\Delta A_{mn}$, edge-symmetry perturbation, $B_{mn}'=B_{mn}+\Delta B_{mn}$, and vertex perturbation, $E_{m}'=E_{m}+\Delta E_{m}$. For simplicity, we refer them as A-, B- and E- perturbation in the following text.

If we do perturbation on the edge $e_{mn}$, Eq.~\eqref{eqs:Fmfpt} is reduced to
\begin{equation}
    \pi_k'-\pi_k = (\langle t_{kn}\rangle-\langle t_{km}\rangle)(\Delta{W_{mn}}\pi_n'-\Delta W_{nm}\pi_m')\pi_k,
\end{equation}
then for $k=m$ and $k=n$ we get two equalities:
\begin{equation}
    \pi_m'-\pi_m = \langle t_{kn}\rangle(\Delta_{W_{mn}}\pi_n'-\Delta W_{nm}\pi_m')\pi_k,\quad  \pi_m'-\pi_m = -\langle t_{km}\rangle(\Delta_{W_{mn}}\pi_n'-\Delta W_{nm}\pi_m')\pi_k,
\end{equation}
which allows us to solve for $\pi_m'$ and $\pi_n'$ explicitly, and express the response on state $k$ using the properties of the unperturbed state
\begin{equation}\label{eqs:exact_singel_edge_response}
    \pi_k' -\pi_k= \frac{\pi_k \left(\pi_m \Delta W_{nm}-\pi_n\Delta W_{mn}\right) \left(\langle t_{km}\rangle-\langle t_{kn}\rangle\right)}{\pi_n \Delta W_{mn} t_{nm}+\pi_m\Delta W_{nm} t_{mn}+1}.
\end{equation}

For A-perturbation, we have $\Delta W_{mn} = W_{mn}(e^{-\Delta A_{mn}}-1)$ and $\Delta W_{nm}=0$, which allows us to express the directed curret from $m$ to $n$ after perturbation (Eq.~(3a) in the main text)
\begin{equation}
    \phi_{mn}' = W_{mn}'\pi_n' =  \frac{\phi_{mn}e^{-\Delta A_{mn}}}{1+(e^{-\Delta A_{mn}}-1)\phi_{mn}\langle t_{nm}\rangle}.
\end{equation}

For B-perturbation, we have  $\Delta W_{mn} = W_{mn}(e^{-\Delta B_{mn}}-1)$ and $\Delta W_{nm}= W_{nm}(e^{-\Delta B_{nm}}-1)$, allowing us to solving for the current on edge $e_{mn}$ after perturbation (Eq.~(3b) in the main text)
\begin{equation}
    j_{mn}^{\prime}=\frac{j_{mn}e^{-\Delta B_{mn}}}{1+(e^{-\Delta B_{mn}}-1)(\phi_{mn}\langle t_{nm}\rangle+\phi_{nm}\langle t_{mn}\rangle)}
\end{equation}

For E-perturbation, Eq.~\eqref{eqs:Fmfpt} can be written as 
\begin{equation}\label{eqs:E_perturbation}
    \begin{aligned}
    \pi_k'-\pi_k 
    &= \sum_{l,l\neq m}(\langle t_{km} \rangle - \langle t_{kl} \rangle ) \Delta W_{lm}\pi_m'\pi_k \\
    &= (e^{\Delta E_m}-1)\sum_{l,l\neq m}(\langle t_{km} \rangle - \langle t_{kl} \rangle )  W_{lm}\pi_m'\pi_k\\
    &=\begin{cases}
        (e^{\Delta E_m}-1)\pi_m'\pi_k\quad &k\neq m\\
        (e^{\Delta E_m}-1)\pi_m'(\pi_m-1)&k=m
    \end{cases}
    \end{aligned}
\end{equation}
To reach the third equality, we use the fact that
\begin{equation}
\sum_{l,l\neq m}\langle t_{km}\rangle W_{lm}- \sum_{l,l\neq m}\langle t_{kl}\rangle W_{lm}=- \langle t_{km}\rangle W_{mm}-\sum_{l,l\neq m}\langle t_{kl}\rangle W_{lm}
=-\sum_l \langle t_{kl}\rangle W_{lm}= 1-\delta_{km}/\pi_m,
\end{equation}
where the last equality is the recursive relation of MFPT [Eq.~\eqref{eqs:mfpt_recur}], as discussed in the next section. Here Eq.~\eqref{eqs:E_perturbation} is the Eq.~(2c) in the main text, and when we take $k=m$, we get:
\begin{equation}\label{eqs:E_pert_exact}
    \pi_m'-\pi_m = (e^{\Delta E_m}-1)\pi_m'(\pi_m-1)\Rightarrow \pi_m' = \frac{\pi_m e^{-\Delta E_m}}{1+(e^{-\Delta E_m}-1)\pi_m}.
\end{equation}
Thus, we obtain the Eq.~(4c) in the main text.

\section{Properties of the mean first-passage time (MFPT)}
\subsection{Equalities of MFPTs}
Here we show two equalities that are essential for the following proof of MFPT inequalities:

\paragraph{Kemeny's Constant}
For any two states $m$ and $n$:
\begin{equation}
    \sum_k \pi_k(\langle t_{km}\rangle-\langle t_{kn}\rangle) = 0
\end{equation}

\paragraph{Recursive relation of MFPTs}
For any states $k$ and $m$:
\begin{equation}\label{eqs:mfpt_recur}
    \sum_l \langle t_{kl}\rangle W_{lm} = -1+\delta_{km}/\pi_m
\end{equation}
which follows from the recursive relation for mean first passage times, for $k\neq m$:
\begin{equation}\label{eq:mfpt_orig}
        \langle t_{km}\rangle = -\frac{1}{W_{mm}} - \sum_{l\neq m}\frac{W_{lm}}{W_{mm}} \langle t_{kl}\rangle,
\end{equation}
where $W_{lm}$ represents the transition rate from state $m$ to state $l$. The first term on the right-hand side is the mean waiting time in the initial state $m$, the second term is the weighted average of all MFPTs to state $l$ from all other states $l\neq m$. Multiplying both sides by $W_{mm}$ and using $W_{mm} = -\sum_{l\neq m}W_{lm}$ gives Eq.~\eqref{eqs:mfpt_recur}. For the case $k=m$, and we can expresse the mean first return time $\langle t_{mm}^r\rangle$ as:
\begin{equation}
    \langle t_{mm}^r\rangle = -\frac{1}{W_{mm}}-  \sum_{l\neq m}\frac{W_{lm}}{W_{mm}} \langle t_{ml}\rangle.
\end{equation}
Note that we have $\langle t_{mm}\rangle =0$  by definition so we can replace the $\sum_{l\neq m}$ with $\sum_{l}$ in the above equation. Since the mean first return time can be expressed as $\langle t_{mm}^r\rangle  = -1/(\pi_m W_{mm})$, substituting it into the above equation and multiplying both sides by $W_{mm}$ gives Eq.~\eqref{eqs:mfpt_recur} under $k=m$.

\subsection{Inequality between MFPT and directed traffic [Proof of Eq. (6) in the main text]}

Here we prove the inequality between MFPT and directed traffic we used in the main text [the Eq. (6) there], which states that: For any two states $m$ and $n$:
\begin{equation}
    \max[\phi_{mn},\phi_{nm}](\langle t_{mn}\rangle+\langle t_{nm}\rangle)\leq 1 \label{mfpt_max}
\end{equation}
where \(\phi_{nm}=W_{nm}\pi_m\) is the steady-state directed traffic from $m$ to $n$ and represents the average number of \(m \to n\) transitions per unit time in stationary states. Since the transitions are instantaneous, the frequency of transitions are solely determined by the time interval between two consecutive transitions. Consequently, the average time between two consecutive \(m \to n\) transitions is
\[
\langle t_{\text{tran}} \rangle 
\;=\; 
\frac{1}{\phi_{nm}}.
\]
Note that $t_{\text{tran}}$ in this case can only include one waiting time $\tau_m$ in $m$ before a direct transition from $m$ to $n$.

Let us consider a generic set of paths (realizations) that can be expressed as 
\begin{equation}
    m\rightarrow n\rightarrow \dots\rightarrow m\rightarrow \dots\rightarrow m\rightarrow n,
\end{equation}
where $m\rightarrow n$ denotes the instantaneous transition from $m$ to $n$ without including the waiting time in $m$.  The notation ``$\rightarrow \dots\rightarrow$'' implies possibly visiting any states or dwelling in states (including $n$ and $m$ themselves). In addition, ``$\rightarrow \dots\rightarrow$'' is defined not to include direct transitions $m\rightarrow n$.

By definition, the total time for each realization in this set is
\begin{equation*}
    t_{\text{re}}=t_{n\rightarrow \dots\rightarrow m}+t_{m\rightarrow \dots\rightarrow m\rightarrow n}=t_{\text{tran}}.
\end{equation*}
All possible events between two consecutive $m\rightarrow n$ transitions are incorporated in this path set. Moreover, all possible first passage events from state $n$ to state $m$ (state $m$ to state $n$) are included in $n\rightarrow \dots\rightarrow m$ ($m\rightarrow \dots\rightarrow m\rightarrow n$), which also gives 
\begin{equation}
    t_{n\rightarrow \dots\rightarrow m}\geq t_{mn},\quad t_{m\rightarrow \dots\rightarrow m\rightarrow n}\geq t_{nm}.
\end{equation}
The equalities hold only when there is no state $m$ in the intermediate steps of $n\rightarrow \dots\rightarrow m$ and no state $n$ in $m\rightarrow \dots\rightarrow m$.

Because the path set contains all possible $t_{\text{tran}}$, $t_{nm}$ and $t_{mn}$, we can properly evaluate 
$\langle t_{\text{tran}}\rangle$, $\langle t_{nm}\rangle$ and $\langle t_{mn}\rangle$ by taking average over all paths in the set.

For each path in the set, we have $t_{\text{re}}=t_{\text{tran}}\geq t_{mn} + t_{nm}$. Taking average over all paths yields
\begin{equation*}
    \frac{1}{\phi_{nm}}=\langle t_{\text{tran}} \rangle  \geq \langle t_{mn} \rangle  + \langle t_{nm} \rangle,
\end{equation*}
i.e., $\phi_{nm}\,\bigl(\langle t_{mn}\rangle + \langle t_{nm}\rangle\bigr)\;\le\;1$.
By symmetry, we also have
\[
\phi_{mn}\,\bigl(\langle t_{nm}\rangle + \langle t_{mn}\rangle\bigr)\;\le\;1,
\]
which completes the proof of \eqref{mfpt_max}.
To better illustrate the proof, we consider a two state equilibrium system as a minimal example. The transition rates and steady states distribution are $W_{12}$, $W_{21}$, $\pi_1$ and $\pi_2$, satisfying $\pi_1+\pi_2=1$ and $W_{12}\pi_2=W_{21}\pi_1$. In this minimal case, the time between two $m\rightarrow n$ is solely contributed by the waiting time in state $n$ and state $m$, i.e., 
\begin{equation*}
    \langle t_{1\rightarrow \dots\rightarrow 2}\rangle=\frac{1}{W_{21}}=\langle t_{21}\rangle,\quad \langle t_{1\rightarrow \dots\rightarrow 1\rightarrow 2}\rangle=\frac{1}{W_{12}}=\langle t_{12}\rangle.
\end{equation*}
From our proof, this indicates that $\phi_{12}=W_{12}\pi_2=\frac{1}{\langle t_{12}\rangle+\langle t_{21}\rangle}$, so that the inequality \eqref{mfpt_max} becomes an equality in this case.
Using the detailed balance condition, it is easy to check 
\begin{equation}
    W_{12}\pi_2\left(\frac{1}{W_{12}}+\frac{1}{W_{21}}\right)=1.
\end{equation}
does hold, validating our proof.

\subsubsection{{Saturation condition and topological interpretation}\label{sec:saturation_MFPT-traffic}}
{
Several universal bounds in the main text ultimately rely on the MFPT--traffic inequality~\eqref{mfpt_max}, so it is useful to explicitly state when the inequality becomes an equality. 
Consider the directed transition $m\to n$ with stationary traffic $\phi_{nm}=W_{nm}\pi_m$. 
Let $t_{nm}^{\rm edge}$ be the (random) waiting time, starting from state $m$, until the \emph{first direct} $m\to n$ jump occurs. 
Let $t_{nm}$ be the standard first hitting time of state $n$ starting from $m$. 
Since any $m\to n$ jump is one particular way of hitting $n$, we have a pathwise inequality $t_{nm}\le t_{nm}^{\rm edge}$, hence
\begin{equation}
    \langle t_{nm}\rangle\le \langle t_{nm}^{\rm edge}\rangle.
\end{equation}

Now consider two consecutive $m\to n$ jumps along the stationary trajectory. Right after the first $m\to n$ jump, the system is in state $n$. 
Therefore, the waiting time between the two consecutive $m\to n$ jumps can be decomposed as (i) the first passage time from $n$ back to $m$, plus (ii) the waiting time from $m$ to the next direct $m\to n$ jump. Taking stationary averages and using that the mean inter-event time equals the inverse directed-traffic, we obtain
\begin{equation}
\frac{1}{\phi_{nm}}
=\langle t_{mn}\rangle+\langle t_{nm}^{\rm edge}\rangle
\ge \langle t_{mn}\rangle+\langle t_{nm}\rangle,
\end{equation}
which is precisely~\eqref{mfpt_max} for the direction $m\to n$ (and similarly for $n\to m$).

Importantly, \emph{equality} in~\eqref{mfpt_max} holds if and only if 
\begin{equation}
t_{nm}^{\rm edge}=t_{nm}\quad \text{almost surely},
\end{equation}
i.e., the first time the process reaches $n$ (starting from $m$) is always realized by the direct edge $m\to n$. 
Equivalently, if we remove the directed edge $m\to n$ from the transition graph, the state $n$ becomes unreachable from $m$ (the edge is an unavoidable kinetic gateway). 
This includes modular ``two-basin'' networks connected by a single bottleneck edge, as shown in Fig.~\ref{fig:MFPT_traffic_saturation_condition}. Therefore, we can also formulate the saturation condition by comparing the MFPT $\langle t_{nm}\rangle$ on the original network with the MFPT $\langle t_{nm}\rangle^{\setminus (m\to n)}$ on the network with the edge $m\to n$ removed. Defining the saturation indicator $\rho_{nm} := \frac{\langle t_{nm}\rangle}{\langle t_{nm}\rangle^{\setminus (m\to n)}}$, we find the equality holds when $\rho_{mn}\to 0$, as illustrated in Fig.~\ref{fig:MFPT_traffic_saturation_condition}(b).

\begin{figure}[h]
    \centering
    \includegraphics[width=1\textwidth]{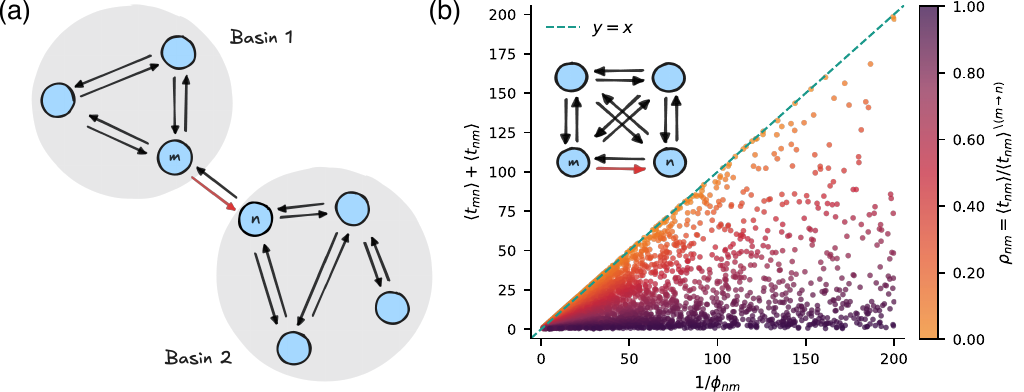}
    \caption{Saturation of the MFPT--traffic inequality~\eqref{mfpt_max}. 
    (a) Schematic networks where the inequality becomes an equality: two subnetworks (basins) connected by a single bottleneck edge. 
    (b) Numerical verification of the saturation condition on  4-state networks with randomly assigned rates. The saturation indicator $\rho_{nm} = \langle t_{nm}\rangle/\langle t_{nm}\rangle^{\setminus ( m\to n)}$ approaches zero in these cases, indicating that all first passages occur via the direct edge.}
    \label{fig:MFPT_traffic_saturation_condition}
\end{figure}

}

\subsection{Triangle inequality for MFPTs [Proof of Eq.~(9) in the main text]} 
For a continuous-time irreducible Markov chain, the mean first passage times satisfy a triangle inequality:
\begin{equation}\label{eq:triangle_inequality_SI}
    \langle t_{kn}\rangle \leq \langle t_{km}\rangle +\langle t_{mn}\rangle,
\end{equation}
where $\langle t_{ab}\rangle$ denotes the mean first passage time from state $b$ to state $a$. The triangle inequality is proven in what follows. 

Define the auxiliary function 
\begin{equation}
    h(l) \equiv \langle t_{kl}\rangle - \langle t_{km}\rangle - \langle t_{ml}\rangle,
\end{equation}
which measures the violation of the triangle inequality. Applying Eq.~\eqref{eqs:mfpt_recur} to any state $n$:
\begin{equation}\label{eq:balance}
\begin{aligned}
    \sum_l W_{ln}\langle t_{kl}\rangle = -1,\quad
    \sum_l W_{ln}\langle t_{ml}\rangle = -1,\quad
    \sum_l W_{ln}\langle t_{km}\rangle = 0,
\end{aligned}
\end{equation}
where the last equation follows from $\langle t_{km}\rangle$ being constant in the sum. Subtracting these equations yields:
\begin{equation}
    \sum_l W_{ln} h(l) = 0,
\end{equation}
By construction, $h(k) < 0$ and $h(m) = 0$. For any state $n$, we can express $h(n)$ as a probability-weighted average:
\begin{equation}
    h(n) = \sum_{l\neq n}\frac{W_{ln}}{|W_{nn}|} h(l).
\end{equation}

Let $n^*$ be a state where $h$ attains its maximum. From the probability-weighted average:
\begin{equation}
\begin{aligned}
    h(n^*) &= \sum_{l\neq n^*}\frac{W_{ln^*}}{|W_{n^*n^*}|} h(l)
    = \frac{W_{kn^*}}{|W_{n^*n^*}|}h(k) + \frac{W_{mn^*}}{|W_{n^*n^*}|}h(m) + \sum_{l\neq k,m,n^*}\frac{W_{ln^*}}{|W_{n^*n^*}|}h(l)\\
    &= \frac{W_{kn^*}}{|W_{n^*n^*}|}h(k) + \sum_{l\neq k,m,n^*}\frac{W_{ln^*}}{|W_{n^*n^*}|}h(l),
\end{aligned}
\end{equation}
where the last equality follows from $h(m) = 0$. Since $h(n^*)$ is the maximum value, $h(l) \leq h(n^*)$ for all $l$, yielding:
\begin{equation}
\begin{aligned}
    h(n^*) \leq \frac{W_{kn^*}}{|W_{n^*n^*}|}h(k) + h(n^*)\sum_{l\neq k,m,n^*}\frac{W_{ln^*}}{|W_{n^*n^*}|}
    = \frac{W_{kn^*}}{|W_{n^*n^*}|}h(k) + h(n^*)\left(1 - \frac{W_{kn^*}}{|W_{n^*n^*}|} - \frac{W_{mn^*}}{|W_{n^*n^*}|}\right),
\end{aligned}
\end{equation}
where we used the probability conservation relation $\sum_{l\neq n^*}\frac{W_{ln^*}}{|W_{n^*n^*}|} = 1$. Rearranging terms:
\begin{equation}
    h(n^*)\left(\frac{W_{kn^*}}{|W_{n^*n^*}|} + \frac{W_{mn^*}}{|W_{n^*n^*}|}\right) \leq \frac{W_{kn^*}}{|W_{n^*n^*}|}h(k).
\end{equation}
Since $h(k) < 0$ and all transition rates are non-negative, the right-hand side is strictly negative while the coefficient of $h(n^*)$ is strictly positive. Therefore, $h(n^*) \leq 0$. As $n^*$ maximizes $h$, this implies $h(n) \leq 0$ for all $n$, completing the proof of the triangle inequality.

\subsubsection{{Saturation condition of triangle inequality}}
{
When Eq.~\eqref{eq:triangle_inequality_SI} is saturated, i.e., $\langle t_{kn}\rangle = \langle t_{km}\rangle + \langle t_{mn}\rangle$, every first-passage path from state $n$ to state $k$ must pass through the intermediate state $m$. In other words, the first passage from $n$ to $k$ decomposes into a first passage from $n$ to $m$ followed by a first passage from $m$ to $k$. Topologically, saturation implies that no route from $n$ to $k$ can bypass $m$, as illustrated in Fig.~\ref{figSI:triangle_inequality_saturation}.

To quantify proximity to saturation, we compare the MFPT $\langle t_{kn}\rangle$ on the original network with the MFPT $\langle t_{kn}\rangle^{\setminus m}$ on the network in which state $m$ has been removed (in practice, all exit rates from state $m$ are set to zero for numerical stability). We define the saturation indicator
\begin{equation}
    \rho_{kn}^{\setminus m} := \frac{\langle t_{kn}\rangle}{\langle t_{kn}\rangle^{\setminus m}}.
\end{equation}
When $m$ is an obligatory waypoint, removing it makes the target $k$ unreachable from $n$, so $\langle t_{kn}\rangle^{\setminus m}\to\infty$ and $\rho_{kn}^{\setminus m}\to 0$. Thus, the triangle inequality becomes an equality precisely when $\rho_{kn}^{\setminus m}\to 0$.

\begin{figure}[h]
    \centering
    \includegraphics[width=1\textwidth]{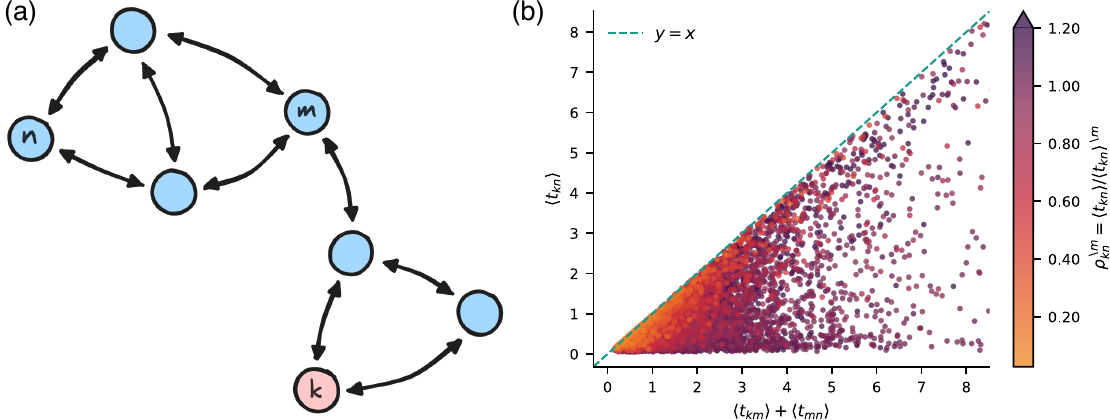}
    \caption{Saturation of the triangle inequality~\eqref{eq:triangle_inequality_SI}. 
    (a) Schematic network topologies where the inequality becomes an equality: two subnetworks (basins) connected by a single bottleneck edge. 
    (b) Numerical verification on randomly generated 6-state networks. The saturation indicator $\rho_{kn}^{\setminus m} = \langle t_{kn}\rangle/\langle t_{kn}\rangle^{\setminus m}$ approaches zero in these cases, confirming that all first-passage paths from $n$ to $k$ pass through the intermediate state $m$.}
    \label{figSI:triangle_inequality_saturation}
\end{figure}
}
\section{Proof of the bound on scaling factor $R_X$ [Eq.~(7) in the main text]}
Here we prove the bound on scaling factor $R_X$ [Eq.~(7) in the main text]. For all three types of perturbations in the main text, we can write the scaling factor as
\begin{equation}
\begin{aligned}
&R_{A_{mn}}:=\frac{\phi'_{mn}}{\phi_{mn}} = \frac{e^{-\Delta A_{mn}}}{1+(e^{-\Delta A_{mn}}-1)\phi_{mn}\langle t_{nm}\rangle},\\
&R_{B_{mn}}:=\frac{j_{mn}'}{j_{mn}}=\frac{e^{-\Delta B_{mn}}}{1+(e^{-\Delta B_{mn}}-1)(\phi_{mn}\langle t_{nm}\rangle+\phi_{nm}\langle t_{mn}\rangle)},\\
&R_{E_m}:=\frac{\pi_m'}{\pi_m}=\frac{e^{-\Delta E_{m}}}{1+(e^{-\Delta E_{m}}-1)\pi_m}.
\end{aligned}
\end{equation}
All of them have the form 
\begin{equation}
    R_X=\frac{e^{-\Delta X}}{1+(e^{-\Delta X}-1)\alpha},
\end{equation}
and for all three type of peturbations, we have $\alpha\in [0,1]$ -- for A- and B- perturbations, $\alpha$ is bounded by the MFPT-flux inequality Eq.~\eqref{mfpt_max}, and for E-perturbation, $\alpha$ is bounded by the steady-state probability $\pi_m\in [0,1]$. Thus we get 
\begin{equation}
    \min\{e^{-\Delta X},1\}\leq R_X\leq \max\{e^{-\Delta X},1\}.
\end{equation}

\subsection{{Saturation of the scaling-factor bound}}
{
The bound above is sharp in the sense that it is attained at the boundary values of the single parameter $\alpha\in[0,1]$ that controls $R_X$:
\begin{equation}
    R_X=\frac{e^{-\Delta X}}{1+(e^{-\Delta X}-1)\alpha}.
\end{equation}
For $\Delta X>0$ (so $e^{-\Delta X}<1$), $R_X$ is monotone increasing in $\alpha$, hence
\begin{equation}
    R_X=e^{-\Delta X}\ \text{iff}\ \alpha=0,
    \qquad
    R_X=1\ \text{iff}\ \alpha=1,
\end{equation}
and the roles of $\alpha=0$ and $\alpha=1$ are exchanged when $\Delta X<0$. 
Therefore the scaling-factor bound in the main text becomes an equality exactly when $\alpha$ is at its extremal values. For the three perturbations considered:
\begin{equation}
\alpha=
\begin{cases}
\phi_{mn}\langle t_{nm}\rangle, & X=A_{mn},\\
\phi_{mn}\langle t_{nm}\rangle+\phi_{nm}\langle t_{mn}\rangle, & X=B_{mn},\\
\pi_m, & X=E_m.
\end{cases}
\end{equation}

In a finite irreducible chain, $\alpha=0$ or $\alpha=1$ can only be reached in singular limits (some rates $\to 0$ or $\to\infty$). 
However, both limits can be approached arbitrarily closely. 
For single edge perturbation, physically, $\alpha\to 1$ corresponds to an \emph{bottleneck} situation in which the perturbed edge is an unavoidable kinetic gateway that dominates the relevant transitions (see the saturation discussion of Eq.~\eqref{mfpt_max} in Sec.~\ref{sec:saturation_MFPT-traffic}); 
$\alpha\to 0$ corresponds to the opposite regime where the perturbed transition carries negligible weight compared to the alternative kinetic pathways. 
Since the nonlinear response satisfies $\Delta_X\langle \mathcal{O}\rangle=(e^{\Delta X}-1)R_X\,\partial_X\langle \mathcal{O}\rangle$, the same extremal conditions control the saturation of the nonlinear-vs-linear response bounds [Eq.~(8) in the main text].
}

\section{Details of the localization principle and range bounds for relative responses \label{sec:localization_deriv}}
\subsection{Localization and bound on range of response for edge-perturbations}
Here, we provide more details on the derivation of the localization principle. First, recall that the inequality
\begin{equation}
    \min_i\frac{x_i}{y_i}\leq\frac{\sum_ix_i}{\sum_iy_i}\leq \max_i\frac{x_i}{y_i}
\end{equation}
holds for $x\in \mathbb{R}$ and $y_i\geq 0$ (at least one positive $y_i$). Taking $x_i=(\pi_i'-\pi_i) \mathcal{O}_i$ and $y_i=\pi_i \mathcal{O}_i\geq 0$, then the above  inequality yields
\begin{equation}
    \min_i \frac{\pi_i'-\pi_i}{\pi_i}\leq \frac{\langle \mathcal{O}\rangle'-\langle \mathcal{O}\rangle}{\langle \mathcal{O}\rangle} \leq \max_i \frac{\pi_i'-\pi_i}{\pi_i}.
\end{equation}
Notably, once there is at least one $\mathcal{O}_i >0$, the inequality holds. For the perturbation on a single edge $X_{mn}$, 
\begin{equation}
    \frac{\pi_i'-\pi_i}{\pi_i}=(e^{\Delta X_{mn}}-1)(\langle t_{im}\rangle-\langle t_{in}\rangle) f^X,
\end{equation}
where $f_B:=j_{mn}'$ and $f_A:=\phi_{mn}'$. Applying the triangle inequality for MFPTs, we find
\begin{equation}
     \min\left\{ \frac{\pi_m'-\pi_m}{\pi_m}, \frac{\pi_n'-\pi_n}{\pi_n} \right\}\leq \frac{\pi_i'-\pi_i}{\pi_i}\leq \max\left\{ \frac{\pi_m'-\pi_m}{\pi_m}, \frac{\pi_n'-\pi_n}{\pi_n} \right\}.
\end{equation}
This completes the proof of the localization principle stated in the main text:
\begin{equation}
\min\left\{ \frac{\pi_m'-\pi_m}{\pi_m}, \frac{\pi_n'-\pi_n}{\pi_n} \right\}\leq \frac{\Delta_{X}\langle \mathcal{O}\rangle}{\langle \mathcal{O}\rangle} \leq \max\left\{ \frac{\pi_m'-\pi_m}{\pi_m}, \frac{\pi_n'-\pi_n}{\pi_n} \right\}.
\end{equation}  
Here, the minimum value is non-positive, and the maximum value is non-negative.

\subsection{{Saturation condition of the localization inequalities}}
{
The localization bounds are instances of the elementary convex-combination inequality
\(
\min_i (x_i/y_i)\le (\sum_i x_i)/(\sum_i y_i)\le \max_i(x_i/y_i)
\)
used at the beginning of this section. 
Therefore, the bounds become \emph{equalities} if and only if the observable puts all of its statistical weight on the states that attain the extremal ratio. 

Concretely, with $x_i=(\pi_i'-\pi_i)\mathcal{O}_i$ and $y_i=\pi_i\mathcal{O}_i\ge 0$, the upper bound is attained iff $\mathcal{O}_i\pi_i>0$ only for states $i$ such that $\epsilon_i:=(\pi_i'-\pi_i)/\pi_i$ equals its maximum; similarly, the lower bound is attained iff $\mathcal{O}_i\pi_i>0$ only for states where $\epsilon_i$ equals its minimum. 
For a single-edge perturbation, the triangle inequality implies that the extrema of $\epsilon_i$ are realized at the perturbed endpoints $m$ and $n$, so the simplest saturating observables are endpoint-localized indicators (e.g., $\mathcal{O}=\mathbf{1}_n$ for the upper bound and $\mathcal{O}=\mathbf{1}_m$ for the lower bound). This formalizes the statement in the main text that localization bounds are saturated by observables localized at the perturbed edge.
}

\subsection{Bound on response range for vertex perturbation}
For vertex perturbation, we don't have a localization principle, but we can still derive a range bound for the relative response, directly from Eq.~\eqref{eqs:E_perturbation}. For $\Delta E_{m}>0$, we have
\begin{equation}
    \begin{aligned}
    0\leq &\frac{\pi_k'-\pi_k }{\pi_k}= (e^{\Delta E_m}-1)\pi_m'\leq (e^{\Delta E_m}-1) \quad &k\neq m\\
    0\geq&\frac{\pi_m'-\pi_m }{\pi_m}=\frac{ (1-e^{\Delta E_m})(1-\pi_m)}{e^{\Delta E_m}+(1-e^{\Delta E_m})\pi_m} \geq e^{-\Delta E_{m}}-1 \quad &k=m
\end{aligned}
\end{equation}

\section{Proof of the nonlinear response resolution limit [Eq.~(13) of the main text]}
In the main text, we have shown that there is a universal bound on the response resolution of arbitrary observables $\mathcal{O}$ with respect to the perturbation on $X=\{A_{mn},B_{mn},E_m\}$, reads 
\begin{equation}\label{eqs:response_resolution_bound}
     \frac{|\Delta_X\langle \mathcal{O}\rangle|}{\sqrt{\text{Var}(\mathcal{O})}}\leq \sinh{\frac{\Delta X}{2}}.
\end{equation}
Here, we provide the detailed derivation of the bound. The first step, as shown in the main text, is using the Cauchy-Schwarz inequality to show that 
\begin{equation}\label{eq:cauchy_schwarz}
    \begin{aligned}
        \Delta_X\langle \mathcal{O}\rangle
        &=\sum_i\mathcal{O}_i(\pi_i'-\pi_i)
        =\sum_i(\mathcal{O}_i-\langle\mathcal{O}\rangle)(\pi_i'-\pi_i)
        =\sum_i\left((\mathcal{O}_i-\langle\mathcal{O}\rangle)\sqrt{\pi_i}\right)\left(\sqrt{\pi_i}(\pi_i'/\pi_i-1)\right)\\
        &\leq\underbrace{\sqrt{\sum_i(\mathcal{O}_i-\langle\mathcal{O}\rangle)^2\pi_i}}_{\sqrt{\text{Var}[\mathcal{O}]}}\underbrace{\sqrt{\sum_i \pi_i(\pi_i'/\pi_i-1)^2}}_{\sqrt{\chi^2(\pi'||\pi)}}\\
        \end{aligned}
\end{equation}
Therefore, to prove the bound Eq.~\eqref{eqs:response_resolution_bound}, we need to show that for all three types of perturbations, the $\chi^2$-divergence is bounded as
\begin{equation}
    \chi^2(\pi'||\pi)\leq \sinh(\Delta X/2)^2
\end{equation}
\subsection{A-perturbation}
For the perturbation on single rate by changing $A_{mn}$ to  $ A_{mn}'=A_{mn}+\Delta A_{mn}$, we can write the $\chi^2$-divergence as
\begin{equation}
    \chi^2(\pi'||\pi):=\sum_i(\pi_i'/\pi_i-1)^2\pi_i=\langle \epsilon^2\rangle =\mathrm{Var}[\epsilon]\leq |\epsilon_m\epsilon_n|
\end{equation}
where we define $\epsilon_i\equiv \pi_i'/\pi_i-1$ as the relative response of state $i$. Note that $\langle \epsilon\rangle=\sum_i \pi_i'-\pi_i=0$, the mean-square of $\epsilon$ equals to the variance of $\epsilon$, and can be further bounded by the range of $\epsilon\in[\epsilon_m,\epsilon_n]$ (from the localization principle) thus gives the last inequality of the above equation. 
Therefore we need to show that
\begin{equation}
    |\epsilon_m \epsilon_n|\equiv\left|\frac{\pi_m'}{\pi_m}-1 \right|\left|\frac{\pi_n'}{\pi_n}-1 \right|\leq \sinh^2{\frac{\Delta A_{mn}}{2}} .
\end{equation}

Using the nonlinear response equality Eq. \eqref{eqs:Fmfpt}, we obtain
\begin{equation}
|\epsilon_m \epsilon_n|
=\frac{(e^{\Delta A_{mn}} - 1)^2\,\phi_{mn}^2 \langle t_{mn}\rangle\langle t_{nm}\rangle}{\left[e^{\Delta A_{mn}} + ( 1-e^{\Delta A_{mn}} )\phi_{mn} \langle t_{nm}\rangle\right]^2} \leq \frac{(e^{\Delta A_{mn}} - 1)^2\phi_{mn} \langle t_{nm}\rangle(1-\phi_{mn} \langle t_{nm}\rangle)}{\left[e^{\Delta A_{mn}} + ( 1-e^{\Delta A_{mn}} )\phi_{mn}\langle t_{nm}\rangle\right]^2}
\end{equation}
where we use the MFPT-flux inequality Eq. \eqref{mfpt_max} to get $\phi_{mn}\langle t_{mn}\rangle \leq 1-\phi_{mn}\langle t_{nm}\rangle$ and gives the last inequality. Then letting $a = \phi_{mn} \langle t_{nm} \rangle$, our aim is to find the maximum of the function
\begin{equation}\label{eqs:A_pert_function}
|\epsilon_m \epsilon_n|\leq f(a): = \frac{(1-e^{\Delta A_{mn}})^2\, a(1 - a)}{\left[e^{\Delta A_{mn}} + (1-e^{\Delta A_{mn}} )\, a\right]^2},
\end{equation}
under $a \in [0, 1]$. Let us denote $c = 1-e^{\Delta A_{mn}}$ for brevity. Then
\begin{equation}
f'(a) = \frac{{\rm d}}{{\rm d}a} \left( \frac{c^2 a(1 - a)}{(1-c + c a)^2} \right) = \frac{c^2 (a (c-2)-c+1)}{((a-1) c+1)^3}.
\end{equation}
Set $f'(a) = 0$, we find the maximum occurs at $a=(c-1 )/(c-2)$, substituting it back to $f(a)$ gives
\begin{equation}
f_{\max}=f\left(\frac{c-1}{c-2}\right)
= \frac{c^2}{4 - 4 c}
\end{equation}
Recall $c =1- e^{\Delta A_{mn}} $, so
\begin{equation}
f_{\max} = \frac{(e^{\Delta A_{mn}} - 1)^2}{4 e^{\Delta A_{mn}}} = \sinh^2 \left( \frac{\Delta A_{mn}}{2} \right).
\end{equation}
Hence we conclude for A-perturbation, the $\chi^2$-divergence is bounded as
\begin{equation}
    \chi^2(\pi'||\pi)\leq |\epsilon_m\epsilon_n|
\le \sinh^2 \left( \frac{\Delta A_{mn}}{2} \right).
\end{equation}
Combined with Eq.~\eqref{eq:cauchy_schwarz}, we get the bound on the response resolution for A-perturbation.

\subsection{B-perturbation}
We next consider the symmetric edge perturbation, $B_{mn}\to B_{mn}+\Delta X$ and $B_{nm}\to B_{nm}+\Delta X$. The localization principle for B-perturbations also holds, thus we can follow the proof of the A-perturbation to get the bound on the $\chi^2$-divergence by the local response on the two endpoints of the perturbed edge, $\chi^2(\pi'||\pi)\leq |\epsilon_m\epsilon_n|$. Then we need to derive the bound on $|\epsilon_m\epsilon_n|$. Using the nonlinear response equality Eq. \eqref{eqs:Fmfpt}, we obtain
\begin{equation}\label{eqs:B-per_range}
    |\epsilon_m\epsilon_n|
    =\frac{(1-e^{\Delta X})^2 j_{mn}^2 \langle t_{mn}\rangle\langle t_{nm}\rangle}{\left[e^{\Delta X} + ( 1-e^{\Delta X} )(\phi_{mn} \langle t_{nm}\rangle +\phi_{nm} \langle t_{mn}\rangle)\right]^2}
\end{equation}
Letting $c=1-e^{\Delta X}$, $a=\phi_{mn}(\langle t_{mn}\rangle+\langle t_{nm}\rangle)$, $b= \phi_{nm}(\langle t_{mn}\rangle+\langle t_{nm})\rangle)$ and $u=\langle t_{mn}\rangle/(\langle t_{mn}\rangle+\langle t_{nm})\rangle)$, we can rewrite Eq.~\eqref{eqs:B-per_range}
\begin{equation}
|\epsilon_m\epsilon_n| = \frac{c^2(a-b)^2u(1-u)}{[1-c+c(a(1-u)+bu)]^2}\leq \frac{c^2u(1-u)}{[1-c+c(a(1-u)+bu)]^2}\tanh\left(\frac{F^{\max}_{c_{mn}}}{4}\right)^2
\end{equation}
where the inequality comes from $|a-b|=|\partial_{B_{mn}}\ln(\pi_m/\pi_n)|\leq \tanh(F^{\max}_{c_{mn}}/4)$ \cite{owen2020Universal}, where $F^{\max}_{c_{mn}}=\max_{c\owns e_{mn}}F_c$ is the maximum cycle affinity passing through edge $e_{mn}$. Now denoting $f(c,a,b,u) = \frac{c^2(a-b)^2u(1-u)}{[1-c+c(a(1-u)+bu)]^2}$, and assuming $a>b$, we can find the maximum is obtained when $u = \frac{1-c}{2-c} $ and $a=1$, $b=0$ (if $a<b$, then $a=0$ and $b=1$). Leading to $\max_{a,b,u}f(c,a,b,u)=c^2/(4-c)=\sinh(\Delta X/2)^2$. Therefore, for B-perturbation, we get a tight bound on the response resolution incorporating thermodynamic information:
\begin{equation}\label{eqs:B_resolution_bound}
    \frac{\Delta_{B_{mn}}\langle \mathcal{O}\rangle}{\sqrt{{\rm Var}[\mathcal{O}]}}\leq \chi^2(\pi'||\pi)\leq \sinh\left(\frac{\Delta B_{mn}}{2} \right)\tanh\left(\frac{F^{\max}_{c_{mn}}}{4}\right)
\end{equation}

\begin{figure}[!hbt]
    \centering
    \includegraphics[width=1\columnwidth]{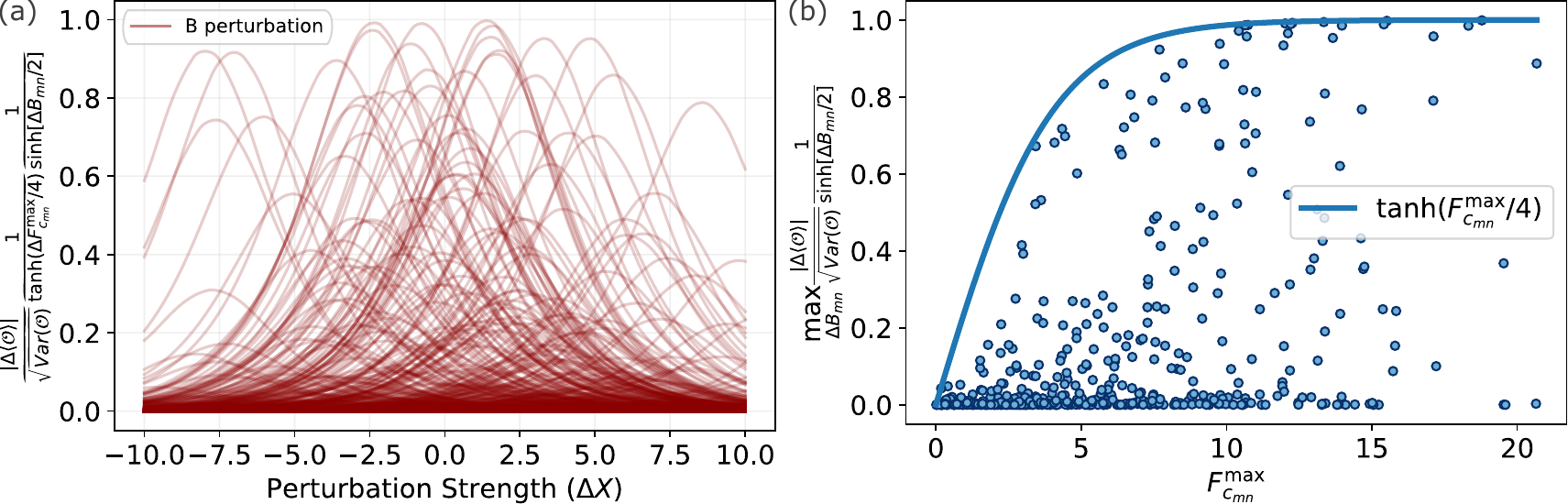}
    \caption{\textbf{Verification of the thermodynamic bound on nonlinear response resolution [Eq~\eqref{eqs:B_resolution_bound}].} (a) The response resolution, $|\Delta_B\langle\mathcal{O}\rangle|/\sqrt{\text{Var}(\mathcal{O})}|$, normalized by its full theoretical upper bound. Data points from randomly generated networks confirm the bound's validity by remaining below 1 for all perturbation strengths $\Delta B_{mn}$. (b) Isolating the thermodynamic contribution. The maximum response resolution, normalized by the perturbation-strength bound $|\sinh(\Delta B_{mn}/2)|$, is plotted against the maximum cycle affinity $F^{\text{max}}_{e_{mn}}$. The data are tightly constrained by the theoretical thermodynamic bound $\tanh(F_{\text{max}}/4)$ (solid line).The numerical data in both panels are generated from simulations of randomly constructed three-state networks.}
    \label{fig:regulation}
\end{figure}

\subsection{E-perturbation}

If we do the vertex peturbation by $E_m \to E_m +\Delta E_m$, we can express the $\chi^2$-divergence using Eq.~\eqref{eqs:E_perturbation} and Eq.~\eqref{eqs:E_pert_exact} as
\begin{equation}
    \begin{aligned}
    \chi^2(\pi'||\pi)
    &=\sum_i(\pi_i'/\pi_i-1)^2\pi_i=(1-e^{\Delta E_m})^2\left(\pi_m (\pi_m'-\pi_m'/\pi_m)^2+\sum_{i\neq m}\pi_i\pi_m'^2\right)\\
    &=(1-e^{\Delta E_m})^2\pi_m'^2(1/\pi_m-1) \\
    &=\frac{(1-e^{\Delta E_m})^2\pi_m(1-\pi_m)}{(e^{\Delta E_m}+(1-e^{\Delta E_m})\pi_m)^2},\\
    &\leq\sinh(\Delta E_m/2)^2
    \end{aligned}
\end{equation}
which takes the form of Eq.~\eqref{eqs:A_pert_function} in the main text and thus we can get the last inequality by the same way as the A-perturbation.

\subsection{{Saturation conditions and physical meaning of the response-resolution bound}}
{
Equation~\eqref{eqs:response_resolution_bound} follows from the chain of inequalities
\begin{equation}\label{eq:RR_chain_SI}
\frac{|\Delta_X\langle\mathcal{O}\rangle|}{\sqrt{\mathrm{Var}[\mathcal{O}]}}
\le \sqrt{\chi^2(\pi'||\pi)}
\le \left|\sinh\!\left(\frac{\Delta X}{2}\right)\right|.
\end{equation}
The first step is the Cauchy--Schwarz inequality in Eq.~\eqref{eq:cauchy_schwarz}, while the second step is the perturbation-strength bound on the $\chi^2$-divergence established above. 
This decomposition makes it transparent what controls \emph{tightness}: (i) the choice of observable through a correlation coefficient, and (ii) the intrinsic distinguishability of the two steady states through $\chi^2(\pi'||\pi)$.

\paragraph{(i) Cauchy--Schwarz saturation and the optimal observable.}
Equality in the first step of Eq.~\eqref{eq:RR_chain_SI} holds if and only if the two vectors in Eq.~\eqref{eq:cauchy_schwarz} are proportional, i.e.,
\begin{equation}\label{eq:CS_sat_cond_SI}
\mathcal{O}_i-\langle\mathcal{O}\rangle = \lambda\,\epsilon_i
\qquad \forall\, i \ \text{with}\ \pi_i>0,
\end{equation}
for some constant $\lambda$ (equivalently, $\mathcal{O}=c+\lambda\,\epsilon$ with an arbitrary constant $c$). 
Consequently, among all observables with a given variance, the maximal achievable nonlinear response resolution is
\begin{equation}\label{eq:opt_resolution_SI}
\max_{\mathcal{O}}\frac{|\Delta_X\langle\mathcal{O}\rangle|}{\sqrt{\mathrm{Var}[\mathcal{O}]}}
=\sqrt{\chi^2(\pi'||\pi)},
\end{equation}
attained by the ``optimal readout'' $\mathcal{O}_{\rm opt}\propto \epsilon$. 
For an arbitrary fixed observable, the tightness of the Cauchy--Schwarz step is the (absolute) Pearson correlation coefficient:
\begin{equation}\label{eq:Corr_decomp_SI}
\frac{|\Delta_X\langle\mathcal{O}\rangle|}{\sqrt{\mathrm{Var}[\mathcal{O}]}\,\sqrt{\chi^2(\pi'||\pi)}}
=\left|\mathrm{Corr}_\pi(\mathcal{O},\epsilon)\right|\le 1.
\end{equation}
Here $\mathrm{Corr}_\pi(\mathcal{O},\epsilon):=\frac{\sum_i \pi_i(\mathcal{O}_i-\langle\mathcal{O}\rangle)\epsilon_i}{\sqrt{\mathrm{Var}[\mathcal{O}]\,\mathrm{Var}[\epsilon]}}$ denotes the Pearson correlation under the unperturbed stationary measure $\pi$.

Thus slack in Eq.~\eqref{eqs:response_resolution_bound} can originate either from a non-optimal observable (small correlation with $\epsilon$), or from a small $\chi^2$--divergence (the perturbed steady state is intrinsically hard to distinguish from the unperturbed one).

\paragraph{(ii) Saturation of the intermediate variance bound: ``binary'' relative response.}
In the A- and B-perturbation proofs we used $\chi^2(\pi'||\pi)=\mathrm{Var}[\epsilon]\le |\epsilon_m\epsilon_n|$. 
This is an instance of the sharp inequality $\mathrm{Var}[X]\le (x_{\max}-\mu)(\mu-x_{\min})$, specialized to $\mu=\langle \epsilon\rangle=0$ and $x_{\min}=\epsilon_m<0<x_{\max}=\epsilon_n$, which yields $\mathrm{Var}[\epsilon]\le -\epsilon_m\epsilon_n$. 
Equality holds if and only if $\epsilon$ takes \emph{only} the two extremal values almost surely (a two-point distribution). 
For edge perturbations, the localization principle ensures that the extrema are realized at the perturbed endpoints $m,n$, hence saturation requires that, up to negligible stationary weight, all states fall into two groups whose $\epsilon_i$ equal either $\epsilon_m$ or $\epsilon_n$. 
Physically this is the effective two-state / two-basin regime produced by an \emph{unavoidable kinetic gateway} (Sec.~III~B): fast equilibration inside each basin combined with a single bottleneck edge makes the local perturbation primarily shift probability mass between the two basins, yielding a nearly piecewise-constant relative-response field $\epsilon_i$.

\paragraph{(iii) Perturbation-specific saturation conditions for $\sqrt{\chi^2}\le|\sinh(\Delta X/2)|$.}
Saturation of the universal response-resolution limit in Eq.~\eqref{eqs:response_resolution_bound} additionally requires that the remaining optimizations in the A/B/E cases are tight.

\emph{A-perturbation.} 
In the A-perturbation proof, the bound is saturated when (a) the MFPT--traffic inequality~\eqref{mfpt_max} is saturated for the perturbed transition (the edge is a kinetic gateway in the sense of Sec.~III~B), and (b) the single kinetic parameter $\alpha=\phi_{mn}\langle t_{nm}\rangle\in(0,1)$ takes the value
\begin{equation}\label{eq:A_alpha_star_SI}
\alpha=\frac{e^{\Delta A_{mn}}}{1+e^{\Delta A_{mn}}}=\frac{1}{1+e^{-\Delta A_{mn}}}.
\end{equation}
A genuine two-state chain realizes condition (a) exactly, and condition (b) can be enforced by tuning the stationary weights of the two states, hence Eq.~\eqref{eqs:response_resolution_bound} is exactly saturable for any fixed $\Delta A_{mn}$.

\emph{E-perturbation.} 
For E-perturbations, the relative response already takes only two values (state $m$ versus all $i\neq m$), so the ``binary'' condition in (ii) holds automatically. 
The $\chi^2$ bound is saturated when the unperturbed occupancy of the perturbed state satisfies
\begin{equation}\label{eq:E_pi_star_SI}
\pi_m=\frac{e^{\Delta E_m}}{1+e^{\Delta E_m}},
\end{equation}
and the observable is any step-like readout distinguishing $m$ from its complement (e.g., $\mathcal{O}=\mathbf{1}_m$), which also guarantees Eq.~\eqref{eq:CS_sat_cond_SI} up to an affine transformation.

\emph{B-perturbation.} 
For B-perturbations, besides the same gateway requirement, maximal $\chi^2$ requires a strongly directed current across the perturbed edge. 
Using the notation introduced above Eq.~\eqref{eqs:B-per_range}, $T=\langle t_{mn}\rangle+\langle t_{nm}\rangle$ and $a=\phi_{mn}T$, $b=\phi_{nm}T$, this corresponds to $|a-b|\to 1$, i.e., one direction dominates. 
Such strong directionality is only possible when there exists a cycle containing $e_{mn}$ with large affinity $F^{\max}_{c_{mn}}\gg 1$ (so that $\tanh(F^{\max}_{c_{mn}}/4)\to 1$ in Eq.~\eqref{eqs:B_resolution_bound}). 
At finite affinity, the refined thermodynamic bound~\eqref{eqs:B_resolution_bound} is itself sharp and can be approached arbitrarily closely when $|a-b|=\tanh(F^{\max}_{c_{mn}}/4)$ (the edge sits on the affinity-dominant cycle) and the remaining gateway/ratio conditions are met.

\paragraph{Physical meaning of tightness and looseness.}
Equation~\eqref{eqs:response_resolution_bound} bounds the \emph{maximal} steady-state distinguishability that any local perturbation of strength $\Delta X$ can generate. 
Near saturation implies that (i) the perturbation acts on a kinetic gateway, producing an effectively two-state redistribution between basins (large $\chi^2$), and (ii) the measured observable is (approximately) the optimal readout aligned with $\epsilon$ (large correlation in Eq.~\eqref{eq:Corr_decomp_SI}). 
Conversely, substantial slack indicates either redundant parallel pathways (the perturbation is kinetically buffered, yielding small $\chi^2$) and/or an observable that is delocalized or poorly aligned with the induced redistribution.
}

\section{Response theory for asymmetric edge perturbations\label{sec:F_perturbation}}

While the main text focuses on rate parameters (activation energy $A_{mn}$), symmetric edge parameters (barrier heights $B_{mn}$) and vertex parameters ($E_m$), the transition rate parametrization also includes asymmetric edge parameters $F_{mn}=-F_{nm}$ corresponding to driving forces. This section provides a comprehensive treatment of responses to perturbations in these asymmetric parameters, which drive the system away from equilibrium when cycle affinities $F_c = \sum_{e_{ij}\in c}F_{ij}$ are non-zero.

\subsection{Linear response to asymmetric edge perturbations}

For perturbations to the asymmetric edge parameter $F_{mn}$, applying the general response relation from Eq.~(1) of the main text yields:
\begin{equation}
    \frac{\partial \ln\pi_k}{\partial F_{mn}}=(\langle t_{kn}\rangle - \langle t_{km}\rangle)\frac{a_{mn}}{2}, \label{F_response}
\end{equation}
where $a_{mn}=\phi_{mn}+\phi_{nm}$ is the edge activity. This relation reveals that asymmetric perturbations are sensitive to the total transition activity on the edge, unlike symmetric perturbations which depend on the net current $j_{mn}=\phi_{mn}-\phi_{nm}$.

The localization principle for F-perturbations follows directly from the triangle inequality for mean first-passage times:
\begin{equation}
    \frac{\partial \ln\pi_n}{\partial F_{mn}} \leq \frac{\partial \ln\pi_k}{\partial F_{mn}} \leq \frac{\partial \ln\pi_m}{\partial F_{mn}}, \label{F_localization}
\end{equation}
establishing that responses are bounded by the values at the endpoints of the perturbed edge.

The sensitivity range for F-perturbations is characterized by:
\begin{equation}
    \left|\frac{\partial \ln(\pi_m/\pi_n)}{\partial F_{mn}}\right| = (\langle t_{mn}\rangle+\langle t_{nm}\rangle)\frac{a_{mn}}{2}.
\end{equation}
Using the MFPT-traffic inequality from Eq.~(6) of the main text, this yields the universal bound:
\begin{equation}
    \left|\frac{\partial \ln (\pi_m/\pi_n)}{\partial F_{mn}}\right| \leq 1. \label{F_range_bound}
\end{equation}

\subsection{Response precision bounds for F-perturbations}

To characterize fundamental limits on F-perturbation detectability, we employ the Fisher information framework. The Fisher information for F-perturbations can be bounded using the variance bound for zero-mean bounded variables:
\begin{equation}
    I(F_{mn}) = \sum_i \pi_i \left(\frac{\partial \ln\pi_i}{\partial F_{mn}}\right)^2 \leq \frac{1}{4}|\partial_{F_{mn}}\ln(\pi_m/\pi_n)|^2 \leq \frac{1}{4},
\end{equation}
where the first inequality follows from the variance bound for bounded zero-mean variables and the second from Eq.~\eqref{F_range_bound}.

This yields the response precision bound for asymmetric edge perturbations:
\begin{equation}
    \frac{|\partial_{F_{mn}}\langle \mathcal{O}\rangle|}{\sqrt{\text{Var}(\mathcal{O})}}\leq \sqrt{I(F_{mn})} \leq \frac{1}{2}, \label{F_precision_bound}
\end{equation}
demonstrating the universal nature of the $1/2$ limit for single-parameter perturbations, regardless of the specific type.

\subsection{Nonlinear response identities for asymmetric perturbations}

The finite perturbation framework extends naturally to F-perturbations. For strong asymmetric perturbations, the response relation becomes:
\begin{equation}
    \frac{\pi_k^{\prime}-\pi_k}{1-e^{-\Delta F_{mn}/2}}=(\langle t_{kn}\rangle - \langle t_{km}\rangle)\pi_k \left({\phi}_{mn}^{\prime}+e^{\Delta F_{mn}/2}{\phi}_{nm}^{\prime} \right), \label{F_finite}
\end{equation}
where $\Delta F_{mn} = F_{mn}^{\prime} - F_{mn}$ and $\phi_{mn} = W_{mn}\pi_m$ is the directed traffic.

This leads to the key identity linking strong and weak F-perturbation responses:
\begin{equation}
    \frac{\Delta_{F_{mn}}\langle \mathcal{O}\rangle}{1-e^{-\Delta F_{mn}/2}}=R_{F_{mn}}\partial_{F_{mn}} \langle \mathcal{O}\rangle, \label{F_strong_weak}
\end{equation}
where $\Delta_{F_{mn}}\langle \mathcal{O}\rangle := \langle \mathcal{O}\rangle^{\prime} - \langle \mathcal{O}\rangle$ and the scaling factor is:
\begin{equation}
    R_{F_{mn}} = \frac{{\phi}_{mn}^{\prime}+e^{\Delta F_{mn}/2}{\phi}_{nm}^{\prime}}{e^{\Delta F_{mn}}a_{mn}} 
\end{equation}

\subsection{Fluctuation-response relations for strong F-perturbations}

The covariance between observables under finite F-perturbations can be expressed as:
\begin{equation}
    \text{Cov}[\mathcal{O}^{(1)},\mathcal{O}^{(2)}]=\sum_{mn} \frac{a_{mn} \left[\frac{\Delta_{F_{mn}} \langle \mathcal{O}^{(1)}\rangle \Delta_{F_{mn}} \langle \mathcal{O}^{(2)}\rangle}{(1-e^{-\Delta F_{mn}/2})^2}\right]}{\left({\phi}_{mn}^{\prime}+e^{\Delta F_{mn}/2}{\phi}_{nm}^{\prime} \right)^2}, \label{F_covariance}
\end{equation}
where all terms are non-negative, providing natural lower bounds for covariance and establishing strong-perturbation variants of fluctuation-response inequalities.

The asymmetric edge parameter framework complements the symmetric perturbations discussed in the main text, providing a complete characterization of nonequilibrium response theory under arbitrary perturbation strengths.

\section{Generalization of response equalities to multiple heat baths and channels\label{sec:multi-channels}}
Multiple heat baths or more than one reaction channels connecting $m$ with $n$ can be modeled as the decomposition of transition rate as 
\begin{equation}
    W_{nm}=\sum_{\nu}W^{\nu}_{nm},
\end{equation}
where $W^{\nu}_{mn}$ is the transition rate in the $\nu$-th channel/contributed by the $\nu$-th heat bath.

If we consider the cases when only the transition rates in the $\mu$-th channel is perturbed (as in the illustrative example of the main text), $j_{nm}$ and $a_{mn}$ is replaced by the current and activity on that channel:
\begin{subequations}
\begin{align}
    \frac{\partial \ln\pi_k}{\partial B^{\mu}_{mn}}&=(\langle t_{kn}\rangle - \langle t_{km}\rangle)
    j_{nm}^{\mu},\label{response1}\\
     \frac{\partial \ln\pi_k}{\partial F^{\mu}_{mn}}&=(\langle t_{kn}\rangle - \langle t_{km}\rangle)
     \frac{a_{mn}^{\mu}}{2}\label{response2},
\end{align}    
 \end{subequations}
where $j^{\mu}_{nm}=W^{\mu}_{nm}\pi_m-W^{\mu}_{mn}\pi_n$ and $a^{\mu}_{nm}=W^{\mu}_{nm}\pi_m+W^{\mu}_{mn}\pi_n$ are the current and activity at the $\mu$-th channel.
Other formula changes similarly, except for the MFPT-traffic inequality, which is invariant, i.e.,
\begin{equation}
   \langle t_{nm}\rangle+\langle t_{mn}\rangle\leq \frac{1}{\max[\phi_{nm},\phi_{mn}]},
\end{equation}
still holds for the directed traffic $\phi_{nm} = W_{nm}\pi_m$. Here, $W_{nm}=\sum_{\nu}W^{\nu}_{nm}$ is the total transition rate from state $m$ to $n$.

\section{Response bounds for normalized current observables}
In the End Matter of the main text, we analyze the response of steady-state currents and current observables, which relies on activity. Here, we define a new type of normalized measurable current observable and show that their response behaviors are independent of dynamical activity, i.e., the timescale of the system's dynamics. The normalized current observable is defined as
$\langle \mathcal{J}\rangle^{nom}:=\sum_{\substack{m, n \\ m < n}} \mathcal{J}_{mn}(p_{mn}\pi_n-p_{nm}\pi_m)$, where we introduce the next-state probability,
\begin{equation*}
    p_{mn}:=p(x_{1}=m|x_{0}=n)=\frac{W_{mn}}{|W_{nn}|}\ (m\neq n),
\end{equation*}
with $x_{0}$ and $x_1$ representing the states before and after a transition ($p_{mm}:=0$).
The normalized current observable can be interpreted as the steady-state average of an observable $\mathcal{J}$ over the joint probability distribution $p_{mn}\pi_n$: $\langle\mathcal{J}\rangle^{nom}=\sum_{m,n}\mathcal{J}_{mn}(p_{mn}\pi_n)$, with $\mathcal{J}_{mn}=-\mathcal{J}_{nm}$. Thus, to characterize the response behavior $\langle \mathcal{J}\rangle^{nom}$, we only need to analyze the response of $p_{mn}\pi_n$. For simplicity, we take the perturbation on $B_{kl}$ as an illustrative example. Recall that $p_{kl}=\frac{W_{kl}}{|W_{ll}|}$, so $\partial_{B_{kl}}p_{kl}=p_{kl}(-1-p_{kl})$ and $\partial_{B_{lk}}p_{kl}=p_{lk}(-1-p_{lk})$, leading to
\begin{align}
    &\frac{\partial \ln (p_{mn}\pi_n)}{\partial B_{kl}}=-(\langle t_{nl}\rangle-\langle t_{nk}\rangle) j_{kl}, \ (e_{mn}\neq e_{kl}) \nonumber \\ 
    & \frac{\partial \ln (p_{kl}\pi_l)}{\partial B_{kl}}=-1-p_{kl}+j_{kl}\langle t_{lk}\rangle \nonumber\\
    & \frac{\partial \ln (p_{lk}\pi_k)}{\partial B_{kl}}=-1-p_{lk}-j_{kl}\langle t_{kl}\rangle.
\end{align}

The Fisher information  $I(B_{kl})=\sum_{m,n}[\partial_{B_{kl}} \ln (p_{mn}\pi_n) ]^2(p_{mn}\pi_n)$ associated with the probability distribution $p_{mn}\pi_n$ can be regarded as the variance of a zero-mean random variable $\partial_{B_{kl}} \ln (p_{mn}\pi_n)$. The triangle inequality $-\langle t_{lk}\rangle\leq \langle t_{nl}\rangle-\langle t_{nk}\rangle\leq \langle t_{kl}\rangle$ together with the inequality $\phi_{mn}(\langle t_{mn}\rangle+\langle t_{nm}\rangle)\leq1$  assure that the random variable is bounded from above and below, $\partial_{B_{kl}} \ln (p_{mn}\pi_n)\in[-a,b]$, with $b+a \leq |j_{kl}|(\langle t_{kl}\rangle+\langle t_{lk}\rangle)+1+\max(p_{kl},p_{lk})\leq 3$. Then using the idea in the main text, we have
\begin{equation}
    I(B_{kl})\leq \frac{a+b}{2}\leq \frac{3}{2}.
\end{equation}
By the Cramer-Rao bound, the response precision is upper bounded as:
\begin{equation}
    \frac{|\partial_{B_{mn}}\langle\mathcal{J}\rangle^{nom}|}{\sqrt{\text{Var}(\mathcal{J})}}\leq \frac{3}{2},
\end{equation}
which is not related to activity. The steady-state variance is defined as $\text{Var}(\mathcal{J}):=\langle \mathcal{J}^2\rangle^{nom}-(\langle\mathcal{J}\rangle^{nom})^2$, where the average is over $p_{mn}\pi_n$.

\section{Details on responses of currents and current observables to finite perturbation}
{In this section, we provide the derivation of the nonlinear response identities of current observables presented in the End Matter [Eq. (18) of the main text].

\textit{Details of the edge perturbation---} 
First, We consider the single-edge perturbation  $A_{mn}\to A_{mn}+\Delta A_{mn}$ ($A$-type) that affects only the forward rate
\begin{equation}
W_{mn}=e^{-A_{mn}},\qquad W'_{mn}=e^{-(A_{mn}+\Delta A_{mn})}=e^{-\Delta A_{mn}}W_{mn},
\end{equation}
while all other transition rates remain unchanged. We denote the perturbed directed traffic by
$\phi_{kl}:=W_{kl}\pi_l$ and $\phi'_{kl}:=W'_{kl}\pi'_l$, and the probability current by
$j_{kl}:=\phi_{kl}-\phi_{lk}$.

The finite-change formula for the stationary distribution under a single-edge $A_{mn}$ perturbation reads
\begin{equation}\label{eq:piA_finite}
\frac{\pi'_k-\pi_k}{e^{\Delta A_{mn}}-1}
= -\bigl(\langle t_{kn}\rangle-\langle t_{km}\rangle\bigr)\,\pi_k\,\phi'_{mn},
\qquad (\forall k),
\end{equation}
whose infinitesimal limit gives
\begin{equation}\label{eq:piA_linear}
\partial_{A_{mn}}\pi_k
= -\bigl(\langle t_{kn}\rangle-\langle t_{km}\rangle\bigr)\,\pi_k\,\phi_{mn}.
\end{equation}
Combining Eqs.~\eqref{eq:piA_finite}--\eqref{eq:piA_linear}, we obtain the nonlinear--linear identity for probabilities,
\begin{equation}\label{eq:piA_scaling}
\frac{\pi'_k-\pi_k}{e^{\Delta A_{mn}}-1}
=
\frac{\phi'_{mn}}{\phi_{mn}}\;\partial_{A_{mn}}\pi_k
\equiv R_{A_{mn}}\,\partial_{A_{mn}}\pi_k,
\qquad
R_{A_{mn}}:=\frac{\phi'_{mn}}{\phi_{mn}}.
\end{equation}

\medskip
\noindent
\textit{Nonlocal response.}
For any edge $(k,l)\neq(m,n)$, the rate $W_{kl}$ is unperturbed, hence
$\phi'_{kl}-\phi_{kl}=W_{kl}(\pi'_l-\pi_l)$. Dividing by $e^{\Delta A_{mn}}-1$ and using
Eq.~\eqref{eq:piA_scaling} (with $k\to l$), we find
\begin{equation}\label{eq:phi_nonlocal}
\frac{\phi'_{kl}-\phi_{kl}}{e^{\Delta A_{mn}}-1}
=
W_{kl}\frac{\pi'_l-\pi_l}{e^{\Delta A_{mn}}-1}
=
R_{A_{mn}}\,W_{kl}\partial_{A_{mn}}\pi_l
=
R_{A_{mn}}\,\partial_{A_{mn}}\phi_{kl}.
\end{equation}
Therefore, for any nonlocal current $j_{kl}=\phi_{kl}-\phi_{lk}$ with $(k,l)\neq(m,n)$,
\begin{equation}\label{eq:j_nonlocal}
\frac{j'_{kl}-j_{kl}}{e^{\Delta A_{mn}}-1}
=
R_{A_{mn}}\,\partial_{A_{mn}}j_{kl}.
\end{equation}

\medskip
\noindent
\textit{Local response.}
For the perturbed edge, we first write the local current response as
\begin{equation}\label{eq:j_local_decomp}
j'_{mn}-j_{mn}=(\phi'_{mn}-\phi_{mn})-W_{nm}(\pi'_m-\pi_m),
\end{equation}
where the second term is directly determined by Eq.~\eqref{eq:piA_finite} with $k=m$ (since $W_{nm}$ is unchanged).
It thus remains to derive the first term $\phi'_{mn}-\phi_{mn}$.

Starting from $\phi_{mn}=W_{mn}\pi_n$ and using $W'_{mn}=e^{-\Delta A_{mn}}W_{mn}$, we compute
\begin{align}
\frac{\phi'_{mn}-\phi_{mn}}{e^{\Delta A_{mn}}-1}
&=
\frac{W'_{mn}\pi'_n-W_{mn}\pi_n}{e^{\Delta A_{mn}}-1}
=
\frac{(W'_{mn}-W_{mn})\pi'_n+W_{mn}(\pi'_n-\pi_n)}{e^{\Delta A_{mn}}-1}
\nonumber\\
&=
\frac{(1-e^{\Delta A_{mn}})W'_{mn}\pi'_n}{e^{\Delta A_{mn}}-1}
+
W_{mn}\frac{\pi'_n-\pi_n}{e^{\Delta A_{mn}}-1}
\nonumber\\
&=
-\phi'_{mn}
+
W_{mn}\frac{\pi'_n-\pi_n}{e^{\Delta A_{mn}}-1}.
\label{eq:phi_local_step}
\end{align}
Using Eq.~\eqref{eq:piA_finite} with $k=n$ and $\phi_{mn}=W_{mn}\pi_n$, we obtain
\begin{equation}\label{eq:phi_local_finite}
\frac{\phi'_{mn}-\phi_{mn}}{e^{\Delta A_{mn}}-1}
=
-\phi'_{mn}
-
W_{mn}\bigl(\langle t_{nn}\rangle-\langle t_{nm}\rangle\bigr)\pi_n\,\phi'_{mn}.
\end{equation}
On the other hand, the linear response of the same traffic is
\begin{align}
\partial_{A_{mn}}\phi_{mn}
&=
(\partial_{A_{mn}}W_{mn})\pi_n+W_{mn}\partial_{A_{mn}}\pi_n
=
(-W_{mn})\pi_n
-
W_{mn}\bigl(\langle t_{nn}\rangle-\langle t_{nm}\rangle\bigr)\pi_n\,\phi_{mn}
\nonumber\\
&=
-\phi_{mn}
-
W_{mn}\bigl(\langle t_{nn}\rangle-\langle t_{nm}\rangle\bigr)\pi_n\,\phi_{mn}.
\label{eq:phi_local_linear}
\end{align}
Comparing Eqs.~\eqref{eq:phi_local_finite} and \eqref{eq:phi_local_linear} yields the local nonlinear--linear identity
\begin{equation}\label{eq:phi_local_scaling}
\frac{\phi'_{mn}-\phi_{mn}}{e^{\Delta A_{mn}}-1}
=
\frac{\phi'_{mn}}{\phi_{mn}}\;\partial_{A_{mn}}\phi_{mn}
=
R_{A_{mn}}\,\partial_{A_{mn}}\phi_{mn}.
\end{equation}
Together with Eq.~\eqref{eq:j_local_decomp} and Eq.~\eqref{eq:piA_scaling} (for $\pi_m$), we conclude that the local
current also satisfies
\begin{equation}\label{eq:j_local_scaling}
\frac{j'_{mn}-j_{mn}}{e^{\Delta A_{mn}}-1}
=
R_{A_{mn}}\,\partial_{A_{mn}}j_{mn}.
\end{equation}

\medskip
\noindent
\textit{Current observables.}
Eqs.~\eqref{eq:j_nonlocal} and \eqref{eq:j_local_scaling} show that, under a single-edge $A_{mn}$ perturbation, \emph{all}
probability currents share the same scaling factor $R_{A_{mn}}=\phi'_{mn}/\phi_{mn}$ relating finite and infinitesimal
responses. Consequently, any current observable
\begin{equation}
\langle \mathcal{J}\rangle := \sum_{k<l}\mathcal{J}_{kl}\,j_{kl}
\end{equation}
obeys the nonlinear--linear response identity
\begin{equation}\label{eq:Jobs_A_scaling}
\frac{\langle \mathcal{J}\rangle'-\langle \mathcal{J}\rangle}{e^{\Delta A_{mn}}-1}
=
\sum_{k<l}\mathcal{J}_{kl}\frac{j'_{kl}-j_{kl}}{e^{\Delta A_{mn}}-1}
=
R_{A_{mn}}\sum_{k<l}\mathcal{J}_{kl}\,\partial_{A_{mn}}j_{kl}
=
R_{A_{mn}}\,\partial_{A_{mn}}\langle \mathcal{J}\rangle.
\end{equation}

For $B$-type perturbation, the current responses to finite perturbation are analogously given by
\begin{subequations}
    \begin{align}
        \frac{j_{kl}^{'}-j_{kl}}{e^{\Delta B_{mn}}-1}&=[\Delta t_{nm}^l\phi_{lk}-\Delta t_{nm}^k\phi_{kl}]j_{mn}^{'}=\frac{j_{mn}'}{j_{mn}}\partial_{B_{mn}}j_{kl}\equiv R_{B_{mn}}\partial_{B_{mn}}j_{kl},\label{FCRR1}\\
        \frac{j_{mn}^{'}-j_{mn}}{e^{\Delta B_{mn}}-1}&=R_{B_{mn}}\partial_{B_{mn}}j_{mn}\label{FCRR2}.
    \end{align}
\end{subequations}

We omit the straightforward derivation of the nonlocal current response [Eq.~\eqref{FCRR1}], as it is fully determined by the responses of the steady-state probabilities $\pi_k$ and $\pi_l$ (similar to the $A$-type perturbation discussed above). One needs to be more careful for the local response. The nonlinear local response of the probability current is derived as: 
\begin{equation}
 \begin{aligned}
\frac{j'_{mn}-j_{mn}}{e^{\Delta B_{mn}}-1}
&=
\frac{
\left(W'_{mn}\pi'_n - W'_{nm}\pi'_m\right)
-
\left(W_{mn}\pi_n - W_{nm}\pi_m\right)
}{
e^{\Delta B_{mn}}-1
}
\\[4pt]
&=
\frac{W'_{mn}\pi'_n - W_{mn}\pi_n}{e^{\Delta B_{mn}}-1}
-
\frac{W'_{nm}\pi'_m - W_{nm}\pi_m}{e^{\Delta B_{mn}}-1}
\\[6pt]
&=
\frac{
(1-e^{\Delta B_{mn}})\phi'_{mn}
+
W_{mn}(\pi'_n-\pi_n)
}{
e^{\Delta B_{mn}}-1
}
-
\frac{
(1-e^{\Delta B_{mn}})\phi'_{nm}
+
W_{nm}(\pi'_m-\pi_m)
}{
e^{\Delta B_{mn}}-1
}
\\[6pt]
&=
\underbrace{
\frac{(1-e^{\Delta B_{mn}})\phi'_{mn}}{e^{\Delta B_{mn}}-1}
-
\frac{(1-e^{\Delta B_{mn}})\phi'_{nm}}{e^{\Delta B_{mn}}-1}
}_{=\,-j'_{mn}}
+
\frac{W_{mn}(\pi'_n-\pi_n)}{e^{\Delta B_{mn}}-1}
-
\frac{W_{nm}(\pi'_m-\pi_m)}{e^{\Delta B_{mn}}-1}
\\[6pt]
&=-j'_{mn}+
\frac{W_{mn}(\pi'_n-\pi_n)}{e^{\Delta B_{mn}}-1}
-
\frac{W_{nm}(\pi'_m-\pi_m)}{e^{\Delta B_{mn}}-1}\\
&=
\frac{j'_{mn}}{j_{mn}}\left(-j_{mn}+W_{mn}\partial_{B_{mn}}\pi_n-W_{nm}\partial_{B_{mn}}\pi_m\right)\\
&=\frac{j'_{mn}}{j_{mn}}\left((\partial_{B_{mn}}W_{mn})\pi_n-(\partial_{B_{mn}}W_{nm})\pi_m+W_{mn}\partial_{B_{mn}}\pi_n-W_{nm}\partial_{B_{mn}}\pi_m\right)\\
&=
\frac{j'_{mn}}{j_{mn}}
\,\partial_{B_{mn}} j_{mn}.
\end{aligned}
\end{equation}
Here we used $\partial_{B_{mn}}W_{mn}=\partial_{B_{mn}}W_{nm}=-1$. The derivation for $j_{nm}$ proceeds identically, as $j_{nm}=-j_{mn}$.

As in the $A$-type case, the same scaling factor also applies to any current observable
$\langle \mathcal{J}\rangle=\sum_{k<l}\mathcal{J}_{kl}j_{kl}$; we therefore omit the repetition of the derivation.

Taken together with the $A$-type derivation above, we conclude that for a single-edge perturbation, the finite responses of \emph{all} probability currents are related to their corresponding linear responses by a universal, perturbation-dependent scaling factor: $R_{A_{mn}}=\phi'_{mn}/\phi_{mn}$ for $A$-type perturbations and $R_{B_{mn}}=j'_{mn}/j_{mn}$ for $B$-type perturbations. This completes the proof of the edge-perturbation part of Eq.~(18) of the main text.

 
}

\textit{Derivations of the nonlinear response to vertex perturbation---}For vertex perturbation, using the relation $\frac{\pi_k^{E_m^{\prime}}-\pi_k^{E_m}}{e^{\Delta E_m}-1}=\pi_m^{E_m^{\prime}}(\pi_k^{E_m}-\delta_{km})$, we find 
\begin{align}
    &\pi_m^{E_m^{\prime}}=\frac{e^{-\Delta E_m}\pi_m^{E_m}}{(e^{-\Delta E_m}-1)\pi_m^{E_m}+1}\\
    &\pi_k^{E_m^{\prime}}=\frac{\pi_k^{E_m}}{(e^{-\Delta E_m}-1)\pi_m^{E_m}+1}.\quad(k\neq m)
\end{align}
Define $\alpha_m=\frac{1}{(e^{-\Delta E_m}-1)\pi_m+1}$ and let $\pi^{\prime}_i:=\pi_i^{E_m^{\prime}}$ for brevity, the expressions are simplified to $\pi_m^{\prime}=\alpha_me^{-\Delta E_m}\pi_m$ and $\pi_k^{\prime}=\alpha_m\pi_k$ for $k\neq m$. Notice that $W_{kl}^{\prime}=W_{kl}$ for $l \neq m$ and $W_{km}^{\prime}=e^{\Delta E_m}W_{km}$ for any $k$. Thus, all steady-state currents $j_{kl}^{\prime}=W_{kl}^{\prime}\pi_l^{\prime}-W_{lk}^{\prime}\pi_k^{\prime}$ after the perturbation on 
$E_m$, are scaled by a same factor $\alpha_m$ as $j_{kl}^{\prime}=\alpha_mj_{kl}$. Then, we have
\begin{equation}\label{current_finite}
    \frac{j_{kl}^{\prime}-j_{kl}}{e^{\Delta E_m}-1}=\frac{\alpha_m-1}{e^{\Delta E_m}-1}j_{kl}=R_{E_m}\frac{\partial j_{kl}}{\partial E_m},
\end{equation}
where $R_{E_m}:=e^{-\Delta E_m}\alpha_m=\pi_m'/\pi_m$ has been defined in the main text. Recall the definition of the current observable is $\langle \mathcal{J}\rangle=\sum_{k<l}\mathcal{J}_{kl}j_{kl}$, so Eq. \eqref{current_finite} gives rise to 
\begin{equation}
    \frac{\langle \mathcal{J}\rangle^{E_m^{\prime}}-\langle \mathcal{J}\rangle^{E_m}}{e^{\Delta E_m}-1}= R_{E_m} \frac{\partial\langle \mathcal{J}\rangle^{E_m}}{\partial E_m},
\end{equation}
as in the main text. If we consider two current observables $\langle \mathcal{J}_1\rangle$ and $\langle \mathcal{J}_2\rangle$, another interesting corollary of the relation $j_{kl}^{\prime}=\alpha_mj_{kl}$ arises as 
\begin{equation}
    \frac{\langle \mathcal{J}_1\rangle^{E_m^{\prime}}}{\langle \mathcal{J}_2\rangle^{E_m^{\prime}}}=\frac{\langle \mathcal{J}_1\rangle^{E_m}}{\langle \mathcal{J}_2\rangle^{E_m}},
\end{equation}
i.e.,
\begin{equation}
     \partial_{E_m}\left (\frac{\langle \mathcal{J}_1\rangle}{\langle \mathcal{J}_2\rangle}\right)=0,
\end{equation}
which generalizes the main result in \cite{malloryigoshin2020} to arbitrary current observables.

\bibliography{refs}